\documentclass[journal]{IEEEtran}
\ifCLASSINFOpdf
\else
\fi
\usepackage{cite}
\usepackage{amsmath,amssymb,amsfonts}
\usepackage{algorithmic} 
\usepackage{graphicx} 
\usepackage{textcomp} 
\usepackage{xcolor} 
\usepackage{booktabs} 
\usepackage{tabularx} 
\usepackage{algorithm} 
\usepackage{hyperref} 
\usepackage{placeins} 
\usepackage{nicefrac} 
\usepackage[caption=false]{subfig} 
\usepackage{newtxtext,newtxmath}  

\usepackage{fontenc}  

\def\BibTeX{{\rm B\kern-.05em{\sc i\kern-.025em b}\kern-.08em

    T\kern-.1667em\lower.7ex\hbox{E}\kern-.125emX}}

\makeatletter
\newcommand{\linebreakand}{%
  \end{@IEEEauthorhalign}
  \hfill\mbox{}\par
  \mbox{}\hfill\begin{@IEEEauthorhalign}
}
\makeatother

\makeatletter
\newcommand{\removelatexerror}{\let\@latex@error\@gobble}
\makeatother

\begin{document}

\title{A Joint Learning Framework for Bridging Defect Prediction and Interpretation}






\author{Guifang Xu,~Chengbin Feng,~Xingcheng Guo,~Zhiling Zhu,~and Wei Wang
\thanks{\textcolor{blue}{This work was supported in part by the National Natural Science Foundation of China under Grant 62262071.} \textit{(Corresponding author: Wei Wang.)}}
\thanks{Guifang Xu, Xingcheng Guo, Zhiling Zhu, and Wei Wang are with the School of Software, Yunnan University, Kunming, Yunnan, China (e-mail: x07191822@163.com; rxij4vcol3@mail.com; leoniddankt@mail.com; wangwei@ynu.edu.cn).}
\thanks{Chengbin Feng is with the School of Information Systems, University of New South Wales, Sydney, Australia (e-mail: chengbin.feng@unsw.edu.au).}
}

\maketitle
\begin{abstract}
Understanding why defect predictors classify software components as defective or clean is essential for software engineers that helps identify the root causes of defects and develop actionable bug-fixing plans. Existing solutions employ various explainable artificial intelligence (XAI) methods to clarify the decision-making processes of defect predictors. However, these post-hoc explanation techniques have two main limitations: 1. the interpretation results do not accurately reflect the model’s decision logic, and 2. they do not contribute to improving the performance of defect predictors. To address these limitations, we treat defect prediction and its corresponding interpretation as two distinct but closely related tasks, proposing a joint learning framework that trains the predictor and its interpreter simultaneously. The novelty of our approach lies in two main aspects:
1. we design a feedback loop that transfers decision logic from the predictor to the interpreter, ensuring a high degree of conciseness for both components.
2. we incorporate interpretation results as a penalty term in the loss function of the joint learning framework. This not only enhances the accuracy of the predictor but also proposes a stronger constraint on the reliability of the interpreter.
We validate our method against several existing explainable software defect predictors across multiple datasets. The results demonstrate its effectiveness.
The source code of our method is available at:
\noindent \href{https://github.com/BugPredictor/software-defect-prediction.git}{https://github.com/BugPredictor/software-defect-prediction.git}
\end{abstract} 

\section{Introduction}
\IEEEPARstart{S}{oftware} defect prediction (SDP) has attracted significant attention in software engineering for over five decades \cite{1,2}. The primary goal of this research is to identify defective code early in the software development process, helping practitioners prioritize quality assurance tasks, especially when resources are limited or deadlines are tight \cite{55}. 

In recent years, inspired by the success of machine learning (ML) and deep learning (DL) in various disciplines, researchers have introduced different ML/DL algorithms into SDP. Despite significant improvements in the performance of SDP models, many practitioners remain hesitant to integrate SDP techniques into practical applications \cite{57}.
Such a situation is primarily due to the lack of interpretability of these models. In practical scenarios, practitioners often ask: What is the reason for predicting a code fragment as buggy? How do different metrics influence the prediction? These questions underscore an urgent requirement for explainability in SDP \cite{3,4, 5}.

To address the interpretation issue, researchers have proposed various explainable SDP models that integrate different explainable artificial intelligence (XAI) techniques with SDP. Existing explainable SDP models are generally categorized into two groups \cite{murdoch2019definitions}: global and local approaches. 
\textit{Global} approaches focus on explaining the overall behavior of a defect predictor, offering insights into how the model makes decisions across all instances \cite{7}. In contrast, \textit{local}  approaches provide clear evidence for why a given instance is predicted as buggy or clean. Compared to \textit{global} approaches, \textit{local} approaches focus on identifying the root cause of specific defects and provide actionable guidance for practitioners to address key questions, such as why a specific prediction was made and how to mitigate associated risks \cite{YU2024}. 
As a result, local interpretation techniques are generally considered more favorable in practice \cite{3,58}.

However, the task properties of interpretation and defect prediction are quite different, making explainable SDP a challenging problem. On one hand, SDP relies on high-complexity ML/DL models to capture intricate relationships among software metrics and defect patterns. These models are often considered as black boxes, with decision-making processes that are difficult to interpret. On the other hand, XAI methods favor simpler models that provide a clear, human-understandable logical reasoning path. Bridging these two competing objectives—complexity and interpretability—remains a significant challenge.
Most existing solutions adopt a post-hoc explanation strategy, in which an SDP model is first trained and subsequently interpreted by XAI techniques \cite{lime, 10}. Although this strategy offers a degree of interpretability, it presents two main challenges: 1. the interpretations often fail to reliably reflect the model’s actual decision logic, and 2. they do not contribute to improving the predictive performance of defect models.

\begin{figure*}[t]
    \centering
    \captionsetup[subfloat]{labelformat=empty} 
    \subfloat[(a) The interpretations by different sampling techniques]{
    \label{fig_1:LIME_RUS}
        \subfloat[The interpretation for the SMOTE-based predictor]{
            \includegraphics[width=0.4\textwidth, height=0.12\textheight]{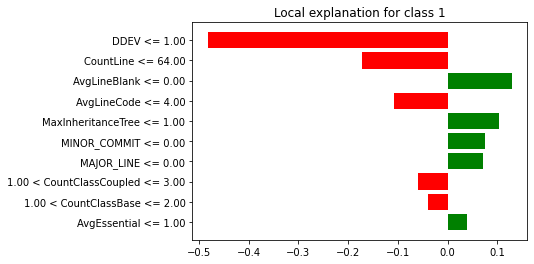}
        }
        \label{fig1:LIME_two}
        \subfloat[The interpretation for the RUS-based predictor]{
            \includegraphics[width=0.4\textwidth, height=0.12\textheight]{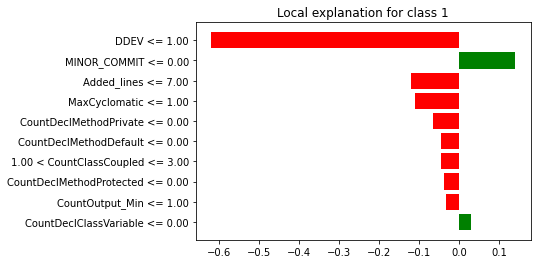}
        }
    }

    \subfloat[(b) Two interpretations produced by LIME]{
    \subfloat[The first interpretation]{
            \includegraphics[width=0.4\textwidth, height=0.12\textheight]{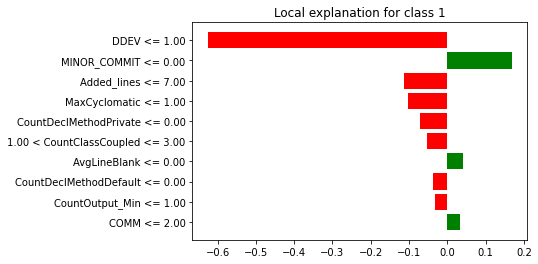}
        }
        \subfloat[The second interpretation]{
            \includegraphics[width=0.4\textwidth, height=0.12\textheight]{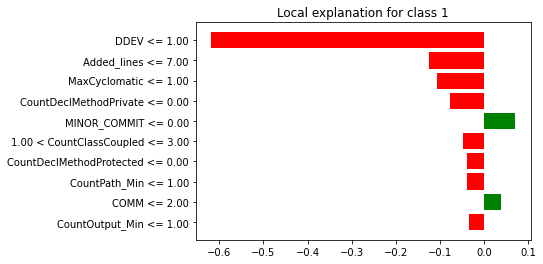}
        }
    }
    \caption{An example of unreliable interpretations.}
    \label{fig_1}
\end{figure*}

To better illustrate the challenges associated with \textit{local} approaches, we examine the interpretation of "\textit{AbstractCommand.java}" from the Apache ActiveMQ project as an example. The explanations generated by LIME using different data sampling techniques (i.e., SMOTE and RUS) are shown in Fig.~\ref{fig_1}~\subref{fig_1:LIME_RUS}, and the explanations generated by executing LIME twice are shown in Fig.~\ref{fig_1}~\subref{fig1:LIME_two}.
The features highlighted in red contribute to the prediction of buggy code, while those highlighted in green contribute to the prediction of clean code.
From the Fig.~\ref{fig_1}~\subref{fig_1:LIME_RUS}, we observe that the explanations vary significantly. Among the ten features selected by SMOTE-based LIME, only two are also selected by RUS-based LIME. Fig.~\ref{fig_1}~\subref{fig1:LIME_two} further illustrates the issue of poor reliability. In the first round of explanation, seven features are identified as correlated with buggy code, whereas in the second round, this number increases to eight.

To address this issue, we treat defect prediction and the corresponding interpretation as two distinct yet closely related tasks, proposing a joint learning framework in this paper. Specifically, we design multiple feedback loops between the interpreter and predictor to simultaneously achieve reliable interpretation and high prediction accuracy. The main contributions are as follows:

\begin{itemize}
    \item To enhance the reliability of interpretation, we introduce two novel loss functions: output fidelity loss and feature fidelity loss. This strategy establishes a knowledge transfer loop between the interpreter and predictor, ensuring both a high level of conciseness in decision-making logic and feature engineering power between these two components.
    \item To improve the accuracy of the predictor, we incorporate the interpretation results as a penalty term in the loss function of the joint learning framework. This strategy enables the predictor to capture more discriminative features guided by the interpretation results while simultaneously imposing stronger constraints on the fidelity of the interpreter to the predictor.
    \item Extensive experiments are conducted on twenty datasets with an in-depth analysis. The results demonstrate that the joint learning framework improves both the reliability of interpretation and the accuracy of the predictor.
\end{itemize}

The remainder of paper is organized as follows: Section II provides an overview of the existing research on SDP and interpretation methods used in SDP. Section III describes our proposed method in detail, while Section IV presents the experimental setup and results. Section V discusses the threats to validity. Finally, Section VI  presents our conclusions and outlines directions for future work.

\section{Related Work}
\subsection{Software defect prediction}\label{subsec_2}
Constructing an SDP model involves three steps. The first step is to transform software metrics, such as cohesion, coupling, complexity, into numerical representations \cite{12}. The second step labels code fragments as buggy or clean based on post-release defects. Finally, leverage the labeled code fragments to train an ML/DL-based classifier for identifying future defect-prone code.

Over the past fifty years, numerous ML algorithms, including fuzzy self-organizing maps, K-means, SVM, Random Forest, and logistic regression, have been extensively employed in SDP research \cite{20, 21, 22, 23, 24, 25, 26}. However, software metrics often exhibit correlations and redundancies \cite{13}, and traditional ML algorithms struggle to capture the complex dependencies among them \cite{14}. To address this issue, various DL models have been introduced in SDP. One of the earliest explorations, Deeper \cite{27}, employed a deep belief network to identify discriminative information in input data. Experimental results showed that Deeper identified 32.22\% more defects than many traditional ML-based approaches. Inspired by the Deeper, researchers have introduced a sophisticated DL algorithms into SDP to enhance performance and broaden application scenarios. For example, \cite{28,30} leaverage convolutional neural networks (CNNs) to build defect prediction models, achieving better performance than other DL-based approaches. Additionally, Qiao et al. employed a DL model to predict the number of defects \cite{29}. Experimental results showed a 14\% reduction in mean squared error and an 8\% increase in squared correlation coefficient. 

To ensure generalizability, we employ DP-CNN, proposed in \cite{30}, as the predictor and evaluate the accuracy differences before and after integrating it into the joint learning framework.

\subsection{Explainable defect prediction}

Efforts in explainable SDP can be broadly divided into two main categories \cite{murdoch2019definitions}: \textit{global} and \textit{local} approaches. The \textit{global} approaches provide a high-level view of how inputs features influence the model’s predictions. Many ML algorithms (e.g., decision trees, logistic regression) and statistical methods (e.g., ANOVA, variable importance) fall into this category\cite{31,32}.
However, \textit{global} approaches cannot offer detailed interpretations for individual code fragment, and their simple structure often fails to ensure optimal predictive accuracy. As a result, many practical applications have adopted \textit{local} methods instead.

\textit{Local} interpretation methods such as LIME \cite{lime} and BreakDown \cite{34}, primarily focus on interpreting specific code fragments. However, recent studies have reported that local techniques often lack reliability under various conditions \cite{36,37,58}. Firstly, \textit{\textbf{interpretations can be inconsistent} when different data sampling techniques are used or when the same interpreter is executed multiple times.} Local interpretation methods typically employ an interpreter to approximate the predictor’s behavior within a local area of the given instance. The local area is generated by the sampling algorithm. However, the randomness of the sampling process makes it difficult to maintain consistency across local areas—regardless of whether the same or different sampling methods are used \cite{58}. As a result, the same code fragment may yield distinct interpretations\cite{37}.
Secondly, \textit{\textbf{interpretations tend to be oversimplified.}} Local interpretation methods usually employ ML algorithms with simple structure as interpreters. For example, LIME utilizes a linear regression model as the interpreter. The interpretation of SDP is oversimplified as a linear transformation between input metrics and prediction outcomes.
Recent study \cite{37} reported that attempting to comprehend an intricate model by employing a simple model might be overly optimistic. Such interpretations often fail to capture the underlying decision-making logic of the predictor.

\subsection{Knowledge distillation}
Knowledge distillation (KD) is a model compression technique that transfers knowledge from a teacher model (e.g., DL models) to a student model (e.g., a shallow neural network) \cite{38,39}. It is considered a viable solution for enhancing the reliability of interpretation due to the knowledge transfer mechanism between the teacher and student models. For instance, \cite{che2016interpretable} utilized distilled knowledge to identify disease patterns, while \cite{biggs2021model} distilled dynamic pricing knowledge from a complex black-box DL model.
In this paper, we treat the predictor and the interpreter as the teacher and student models, respectively, and design knowledge transfer channels based on the KD principle. The key distinction of our approach lies in the collaborative training mechanism: unlike existing KD methods, which keep the parameters of the teacher model fixed, our method trains both the interpreter and the predictor in a collaborative manner.

\section{Proposed Method}
\begin{figure}[!tb]
\centering
{\includegraphics[width=\columnwidth]{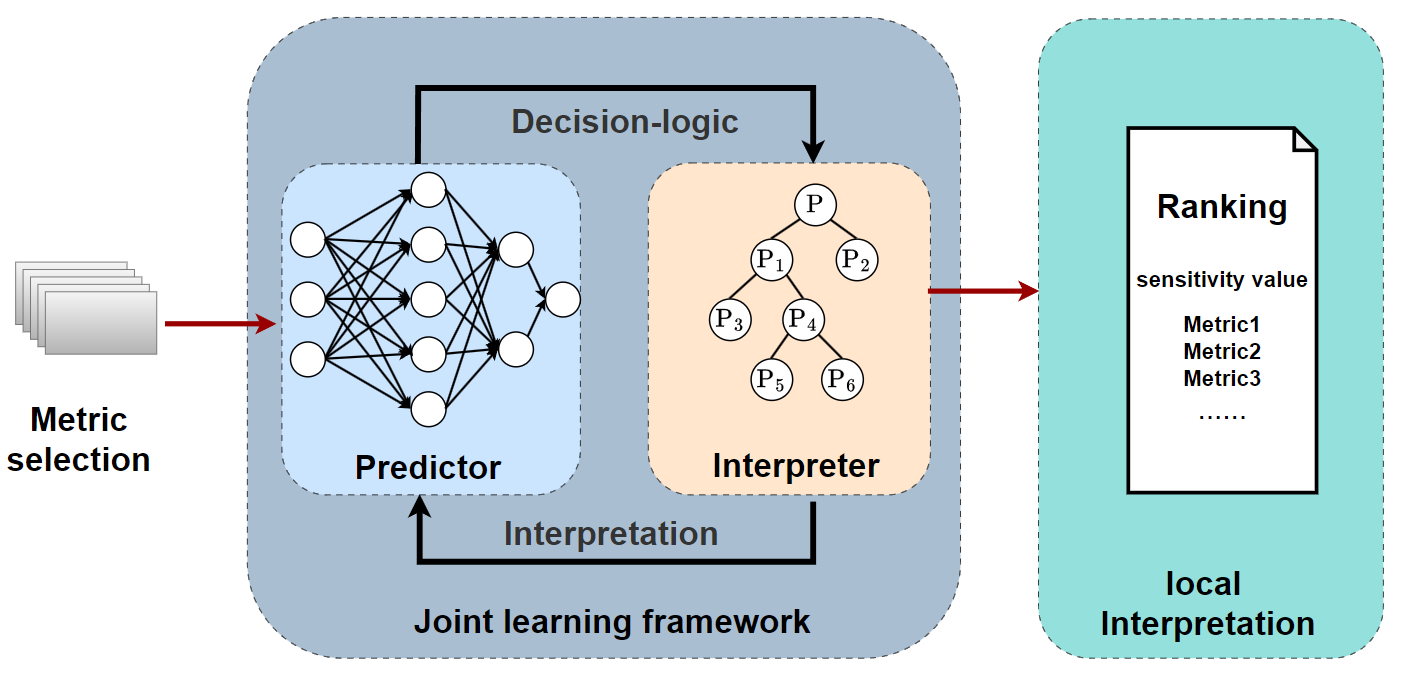}}
\caption{The framework of our approach.}
\label{fig1}
\end{figure}

Fig.~\ref{fig1} provides an overview of our proposed approach, which consists of four modules: metric selection, predictor and interpreter design, joint learning framework, and local interpretation.

\subsection{Metric selection}
We selected sixteen metrics for our method based on a review of existing literature \cite{12,43,44}. The details of these metrics are presented in Table \ref{tab1}. The selection was guided by two criterias: (1) expressiveness: The metrics needed to be sufficiently expressive to capture subtle variations in code quality, structure, and complexity. (2) effectiveness: Each selected metric has been demonstrated its utility in SDP tasks \cite{22,27,45,46}. According to these criteria, we constructed a dataset that encompasses a broad spectrum of code attributes, ranging from complexity measures to process-related characteristics.

\begin{table}[tb]
\centering
\caption{Metrics suite}
\label{tab1}
\begin{tabular}{ll}
\toprule
\multicolumn{1}{c}{\textbf{Feature}} & \multicolumn{1}{c}{\textbf{Description}} \\ 
\midrule
WMC & Weighted methods per class  \\
\midrule
DIT & Depth of Inheritance Tree  \\
\midrule
NOC & Number of Children \\
\midrule
CBO & Coupling between object classes  \\
\midrule
RFC & Response for a Class \\
\midrule
LCOM & Lack of cohesion in methods  \\
\midrule
NPM & Number of Public Methods  \\
\midrule
DAM & Data Access Metric \\
\midrule
MOA & Measure of Aggregation  \\
\midrule
MFA & Measure of Functional Abstraction  \\
\midrule
CAM & Cohesion Among Methods of Class  \\
\midrule
IC & Inheritance Coupling  \\
\midrule
CBM & Coupling Between Methods  \\
\midrule
AMC & Average Method Complexity  \\
\midrule
LOC & Lines of Code  \\
\midrule
CC & McCabe's cyclomatic complexity \\
\bottomrule
\end{tabular}
\end{table}

\subsection{Defect prediction and interpretation model}
In this section, we provide the detailed information about the predictor and interpreter.

\subsubsection{Defect predictor}
\begin{figure}[!t]
\centering
{\includegraphics[width=\columnwidth]{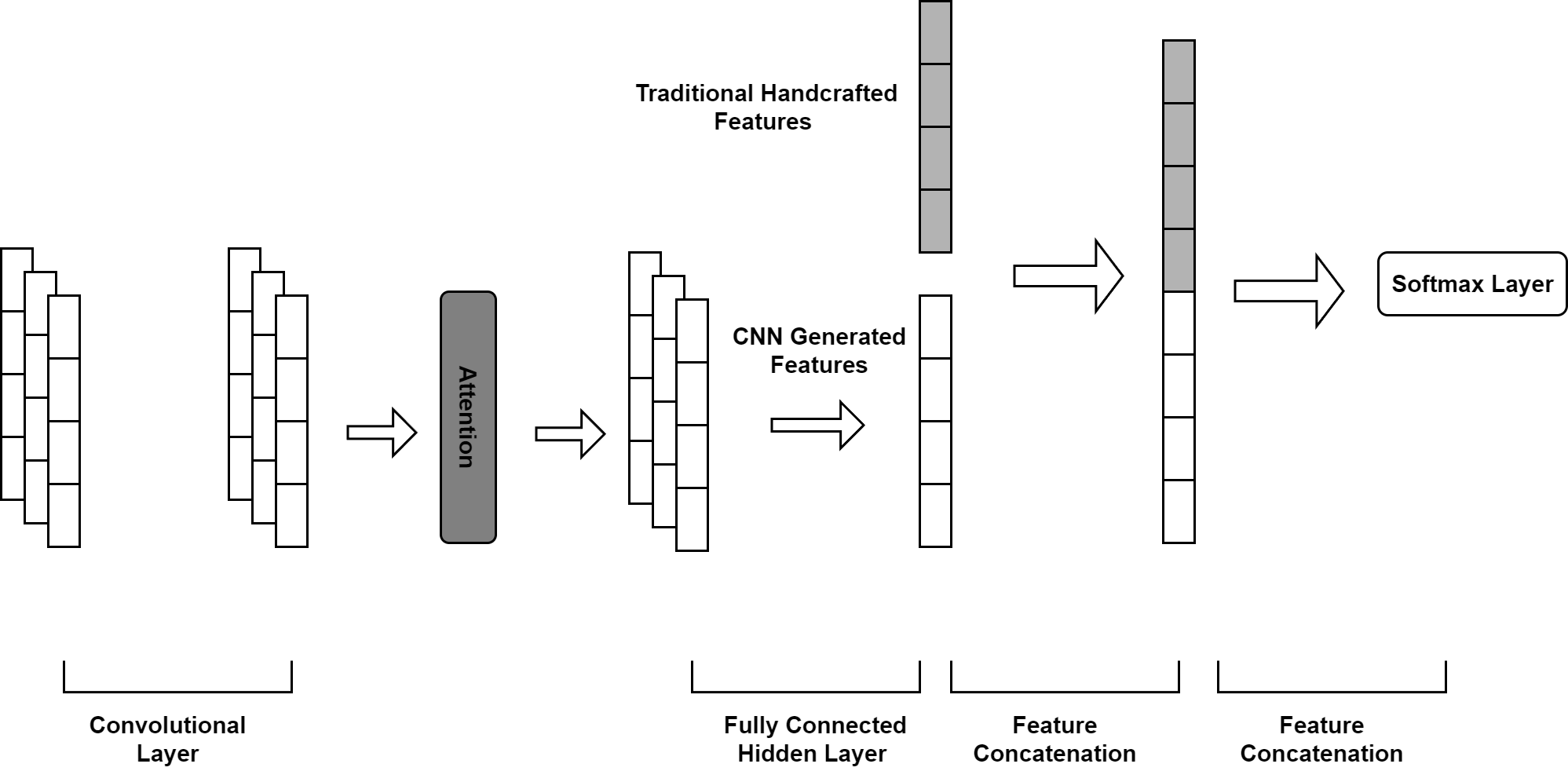}}
\caption{Model Architecture of DP-CNN.}
\label{fig2}
\end{figure}

In this paper, we chose a DL-based model, DP-CNN \cite{30} as the predictor. It consists of three main components: a convolution layer,  a max pooling layer, and  a classification layer. To further improve predictive accuracy, we extend the original model by integrating an attention mechanism. The overall architecture of DP-CNN is illustrated  in Fig.~\ref{fig2}. 

The DP-CNN model predicts defect-proneness through the following process: Selected metrics are first fed into the convolutional layer to capture correlations among features. The output is then passed through the max pooling layer to highlight key features while reducing redundancy. Next, the attention layer focus on critical defect patterns by dynamically assigning weights. Finally, a fully connected network followed by softmax classifier generates the prediction. We choose the CNN-based model as the predictor for three key reasons: (1) Sparse connectivity enables a wider receptive field to capture non-linear correlations. (2) Weight sharing ensures positional invariance in feature extraction \cite{30}. (3) Max pooling enhances robustness by preserving significant features, while the attention layer further refines feature importance.

\subsubsection{Interpretation model}
Following the KD principle, we employ the soft decision tree (SDT) as the interpreter \cite{sivaprasad2023evaluation}. The SDT is a variant of the fuzzy decision tree, sharing a similar structure with the ordinary decision tree but differing in the definition of nodes.
In SDT, nodes are defined as neurons with learnable weights $W$ and biases $b$. The output of each node is defined as $P_i(x_i) = \sigma(x_iW_i + b_i)$, which determines the probability of transitioning to the right subtree. The $\sigma(\cdot)$ denotes the activation function and $x_i$ is the input of the $i$-th node. The output at the $l$-th leaf node is defined as:
\begin{equation}
     Q_k^l=\frac{{\rm exp}(\phi_k^l)}{\sum_k {\rm exp}(\phi_k^l)}
\label{eq1}     
\end{equation}      
where $Q_k^l$ is the probability at the $l$-th leaf of type $k$ defect, and $\phi_k^l$ is the learned feature at that leaf.     

The reasons we use SDT as an interpreter are as follows:
Owing to its neural-like node structure, SDT exhibits comparable feature engineering capabilities to CNN, enabling it to simulate the decision-making knowledge of CNN. Furthermore, the complexity of SDT has been reduced, making the decision-making process more comprehensible.

\subsection{Joint learning framework}
The architecture of the joint learning framework is presented in Fig.~\ref{fig3}, each feedback loop corresponds to a loss function. Given training data set $D=\{\left(x_i,y_i\right)_{i=1}^N\}$, $x_i\in R^n$, $y_i\in \{0,1\}$, the predictor $f$ and interpreter $g$, the loss function for joint learning framework is defined as follows:
\begin{figure}[tb]
\centering
\includegraphics[width=\columnwidth]{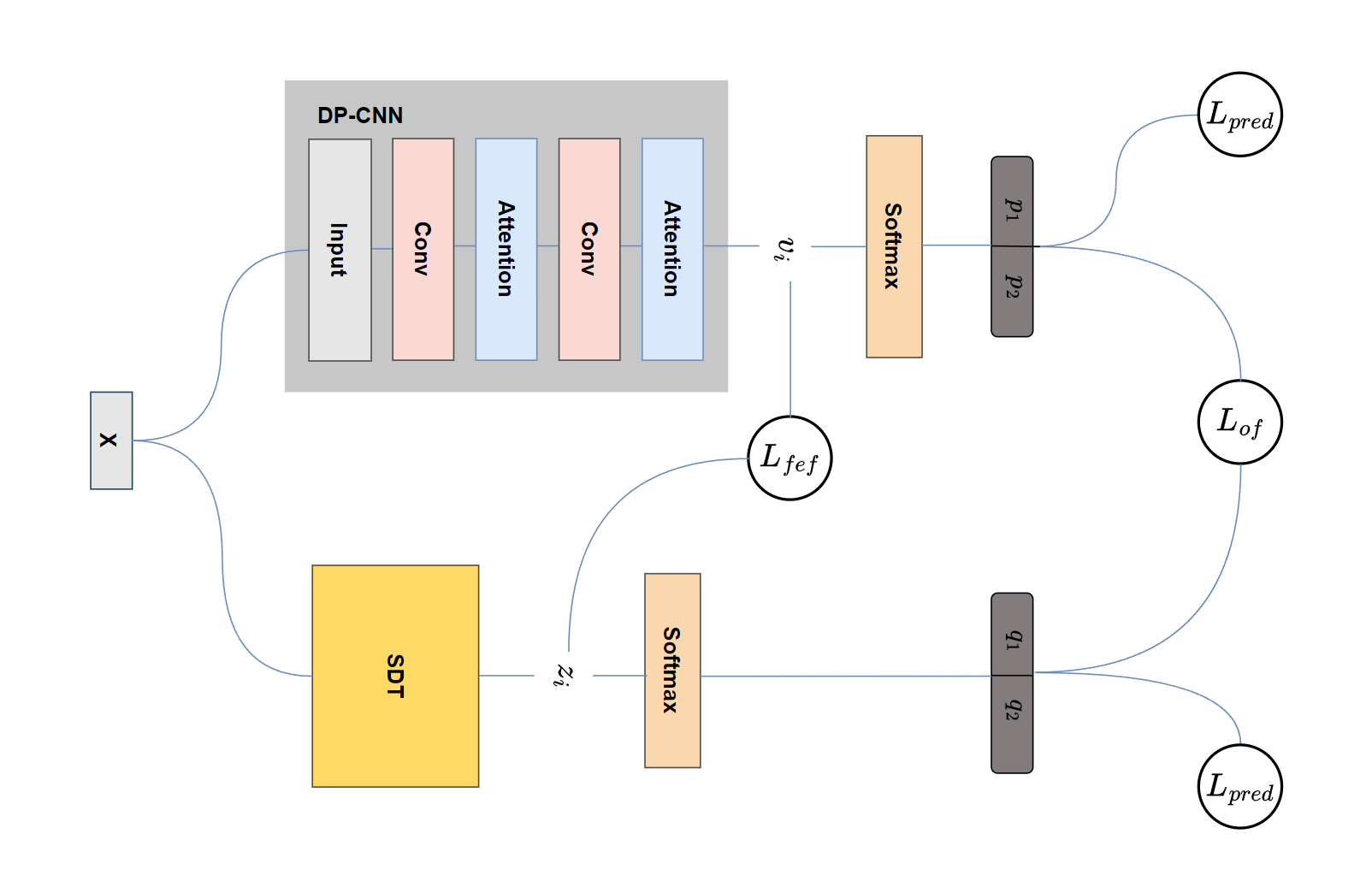}
\caption{The architecture of joint learning framework.}
\label{fig3}
\end{figure}

\begin{equation}
    L=   \alpha L_{pred}(f, D) + \beta  L_{pred}(g, D)+ L_{int}(f,g,D)
\label{eq2}
\end{equation}
where the hyperparameters $\alpha$ and $\beta$ denote the relative importances of the subloss to the the overall loss.

The original loss function of KD consists of two parts: the distillation loss $L_{soft}$ (which is equivalent to $L_{int}(f,g,D)$ in Eq. \eqref{eq2}) and the student loss $L_{hard}$ (which is equivalent to $L_{pred}(g,D)$ in Eq. \eqref{eq2}). $L_{soft}$ quantifies the difference between the teacher model and the student model, which serves as a feedback loop for transferring knowledge from the teacher model to the student model. 
$L_{hard}$ evaluates the difference between the student model’s predictions and the ground truth. 

In this paper, we introduce two key modifications to the original KD loss function: (1) added the output fidelity loss $L_{pred}(f, D)$, and (2) revised the $L_{soft}$.
$L_{pred}(f, D)$ shares a similar definition with $L_{pred}(g, D)$ indicating the consistency of the predictor's output to the ground truth.
\begin{equation}
    L_{pred}(f, D)=\sum_{i=1}^{|D|}\left(f(x_i)-y_i\right)^2
\label{eq4}
\end{equation}
\begin{equation}
    L_{pred}(g, D)=\sum_{i=1}^{|D|}\left(g(x_i)-y_i\right)^2
\label{eq5}
\end{equation}
Introducing output fidelity loss here can enhance both prediction accuracy and interpreter fidelity simultaneously. Within the KD framework, predictor and interpreter are mutually dependent. Minimizing the output fidelity loss will encourage the predictor to achieve heigher accuracy. Moreover, improving the accuracy of the predictor is equivalent to indirectly enforcing the interpreter to provide a more accurate explanation of the predictor's decision logic.
Therefore, incorporating the output fidelity loss can boost the performance of the predictor  but also yield a more robust and insightful knowledge based on the KD principle.

The revised the loss function $L_{soft}$ (denoted as $L_{int}(f,g, D)$ in Eq. \eqref{eq2}) is presented in Eq. \eqref{eq_5}. It involves two parts, $L_{of}(f,g,D)$ and $L_{fef}(f,g,D)$. 
$L_{of}(f,g,D)$ is equivalent to the distillation loss in the original KD framework. The feature fidelity loss $L_{fef}(f,g,D)$ evaluates the similarity between the feature maps generated by predictor and interpreter.
Minimizing $L_{of}(f,g,D)$ and $L_{fef}(f,g,D)$ can enforce the interpreter to achieve similar feature engineering power of predictor.
\begin{equation}
L_{int}(f,g,D) = \lambda L_{of}(f,g,D) + \gamma L_{fef}(f,g,D) 
\label{eq_5}
\end{equation}
where $\lambda$ and $\gamma$ are hyperparameters.
$L_{of}(f,g,D)$ shares a similar definition with $L_{soft}(f,g,D)$. 
\begin{equation}
L_{of}(f,g,T) = -\sum_{i=1}^{|D|} q_i^T \log(p_i^T)
\label{eq6}
\end{equation}
where $ q_i^T = \frac{{\rm exp}(\nicefrac{z_i}{T})}{\sum_{k}^{N} {\rm exp}(\nicefrac{z_k}{T})}$ and $p_i^T = \frac{{\rm exp}(\nicefrac{v_i}{T})}{\sum_{k}^{N} {\rm exp}(\nicefrac{v_k}{T})}$ are the classification probability of instance $x_i$ generated by $g$ and $f$, $z_i$ and $v_i$ are the logits of $x_i$. $T$ is the temperature hyperparameter \cite{sivaprasad2023evaluation}. 
Furthermore, the feature  fidelity loss $L_{fef}$ is defined as follows:
\begin{equation}
L_{fef} = \sum_{i}^{|D|} \left(
d(z_i)-v_i
\right)^2
\label{eq7}
\end{equation}
where function $d(\cdot)$ adjusts $z_i$ to the same dimension of $v_i$.

\begin{figure}[t]
    \centering
    \includegraphics[width=\columnwidth]{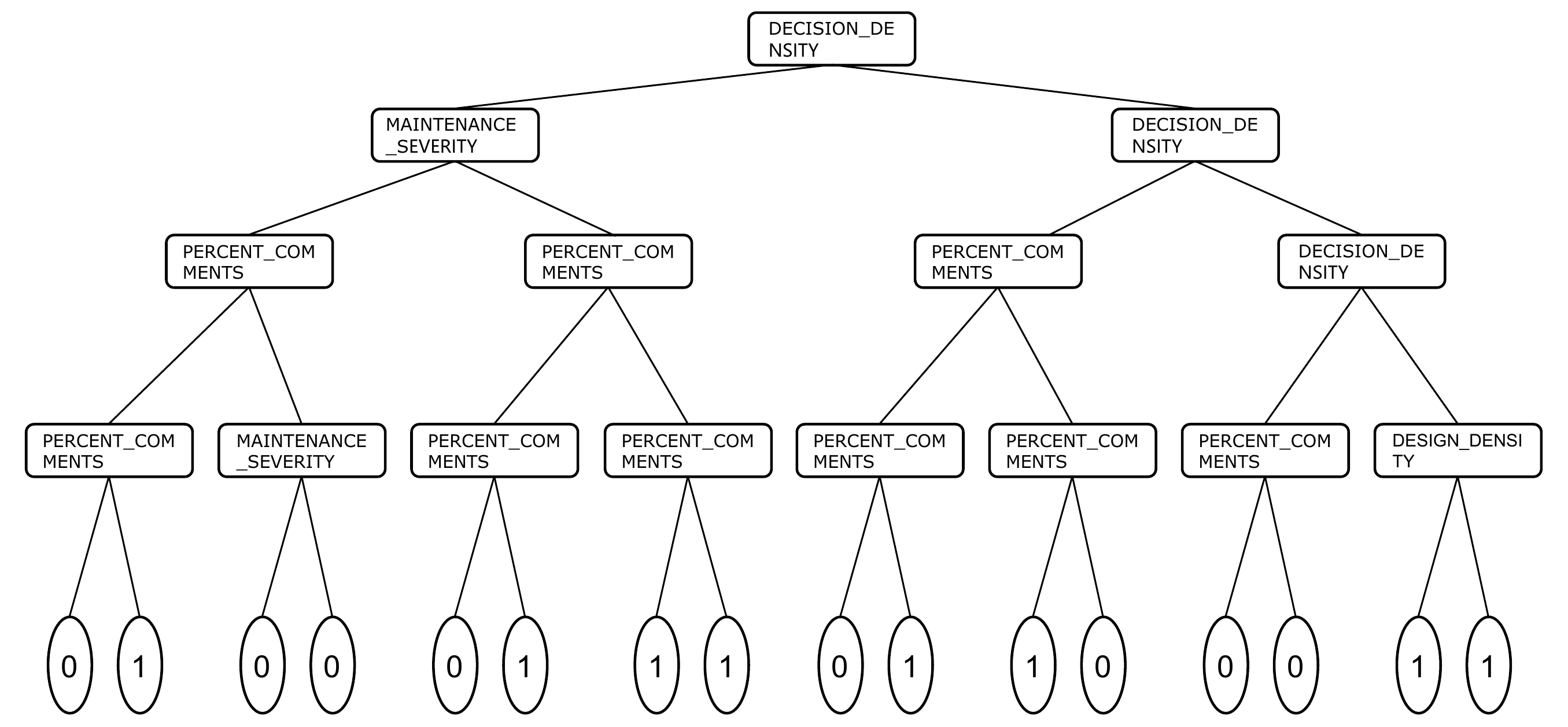}
    \caption{The interpretations of the proposed method on PC1 dataset.}
    \label{PC2_tree}
\end{figure}

To further demonstrate the proposed joint learning framework, we use the PC2 dataset as an example. Each data point in PC2 consists of twenty software metrics and one label. According to Eqs. (3, 4), DP-CNN and SDT iteratively optimize their parameters to minimize discrepancies between predictions and actual labels. Crucially, improving prediction accuracy of DP-CNN inherently enforces the SDT to provide a more precise explanation of the predictor’s decision-making process, which indirectly improve interpreter's reliability. Meanwhile, according to Eq. (6), the features generated by DP-CNN and SDT should be similar, ensuring that both models have comparable decision logic, thereby enhancing the interpretability of SDT. Additionally, Eq. (7) ensures consistency between SDT’s selected features and the input metrics, which strengthens the reliability of the interpreter. As presented in Fig.~\ref{PC2_tree}, the decision logic of DP-CNN is interpretered as a set of ``IF-THEN'' paths in the SDT, which is similar to the decision tree doess \cite{sivaprasad2023evaluation}.


\subsection{Local interpretation}
In this section, we focus on the interpretability for an instance. 
Given a metric set $A=\{a_1,a_2,...a_d\}$, the sensitivity of SDT for an instance $x$ on the $a_i$ is defined as follows.
\begin{equation}
S(a_i) = \frac{\Delta g_i(x)}{\Delta a_i}
\label{eq8}
\end{equation}
$\Delta g_i(x) = \lvert g(x) - g(x + {\rm One}(a_i)\Delta a_i) \rvert$
is the changes of the output of interpreter aroused by the perturbation of the metric $a_i$. $\Delta a_i$ is the standard variance of $a_i$. ${\rm One}(a_i)$ is the onehot encoder of $i$-the metric. 
\begin{equation}
\Delta a_i = \sqrt{\frac{1}{N} \sum_{i=1}^{N} (a_i - \bar{a})^2}
\label{eq9}
\end{equation}
\begin{equation}
\bar{a}_i = \frac{1}{N} \sum_{j}^{N} a_{j}^{i} 
\label{eq10}
\end{equation}
where $N$ is the number of training data, and $a_{j}^{i}$ is the $i$-th metric of $x_j$.
According to Eq. \eqref{eq8}, we interpret the correlation between the instance and predictive result as the sensitivity to feature perturbation. Given an instance $x_i$, if a metric has a greater sensitivity, this means the metric has a stronger influence on the prediction result.

Since the $\Delta a_i$ is a constant and the parameters of interpreter are fixed after training, the sensitivity $S$ can provide a stable and consistent interpretation for any instance. The algorithm for performing local interpretation is presented below:

    \begin{algorithm}[htbp]
        \caption{Local Interpretation}
        \begin{algorithmic}[1]
            \REQUIRE interpreter $g$, instance $x$, metric set $A=\{a_1, a_2, \dots, a_d\}$, and  hyperparameters: $\lambda, \gamma$.
            \ENSURE Sensitivity index $SI$
            \STATE \algorithmicfor\ $a_i \in A$ \algorithmicdo\
                \STATE \quad $\text{Calculate }\bar{a}_i$ \hfill Eq. \eqref{eq10}
                \STATE \quad $\text{Calculate } \Delta a_i $ \hfill Eq. \eqref{eq9}
                \STATE \quad $\text{Calculate }\Delta g_i(x)$
                \STATE \quad $S =  \cup \ S(a_i)$  \hfill Eq. \eqref{eq8}
            \STATE \algorithmicend\ \algorithmicfor\ 
            \STATE $SI \leftarrow$ \text{ Sorts $S$ in descending order}
            \STATE \textbf{return} $SI$
        \end{algorithmic}
    \end{algorithm}

The loop structure (lines 2 to 5 in the algorithm) computes the sensitivity of each metric and
generates the sensitivity set $S$. Line 7 rearranges the sensitivity set $S$ in descending order and puts the index of $S$ to $SI$.

\section{Experiment}\label{sec4}
To objectively evaluate the performance of our method, we conducted an experiment to quantify the performance of the proposed method and the baselines on twenty datasets, aiming to answer the following research questions:

\textbf{RQ1: }What improvements does the proposed method offer for the consistency of interpretations?

As mentioned in Section II, many \textit{local} methods tend to yield inconsistent interpretations, which greatly affects the reliability of explainable SDP. To evaluate the consistency of interpretation, we adopt the Coincidence Degree (CD) in RQ1 \cite{parekh2021framework}. Given an interpreter $g$, the CD of an instance $x_i$ is defined as follows :

\begin{align}
    CD(x_i) =  \frac{1}{|M|} \left| \bigcap_{j=1}^{|M|} g_j(x_i) \right| \label{eq13}
\end{align}
where $\bigcap_{j=1}^{|M|}{g_j(x_i)}$ is the intersection of the multiple interpretations for one instance, $|M|$ represents the number of interpretations or the number of data sampling techniques used. $CD-k\%$ indicates the coincidence of the top $k$ metrics. A greater $CD$ indicates greater consistency.


Furthermore, another issue related to the consistency of interpretation is that the predictor and interpreter may generate different predictive results for the same instance. To evaluate the consistency between predictor and interpreter, we introduce Fidelity of Interpretation (FI) \cite{parekh2021framework,lakkaraju2020robust,bang2021explaining} here. Given dataset $D$, predictor $f$ and interpreter $g$, the fidelity is defined as:

\begin{equation}
    FI=\frac{1}{|D|}\sum_{(x_i,y_i)\in D} \mathbb{I}\left(f(x_i),g(x_i)\right)
\label{eq12}
\end{equation}
$ \mathbb{I}(\cdot)$ is an indicator function, which generates 1 when $f(x_i)=g(x_i)$, 0 otherwise.

In addition to FI, we also argue that the prediction produced by an interpreter should also align with the ground truth. The alignment indicates how much correct decision knowledge is captured from the predictor.
In this paper, we use the Accuracy of Interpreter (AI) to measure this alignment. Given the dataset $D$ and interpreter $g$, we define the accuracy as:
\begin{equation}
    AI=\frac{1}{|D|}\sum_{(x_i,y_i)\in D} \mathbb{I}\left(g(x_i),y_i\right)
    \label{eq111}
\end{equation}
where $\left|D\right|$ denotes the cardinality of $D$, $g\left(x_i\right) $ is the prediction made by the interpreter for the instance $x_i$.

\textbf{RQ2:} What performance improvements does the joint-learning framework achieve?

There are two goals in validating the predictor trained by the joint learning framework (denoted as Joint\_CNN): 
(1) to assess the positive impact of the joint learning framework on predictor's 
performance, and (2) to determine whether its predictive performance outperforms other notable predictors. 
For the first goal, we evaluate model performance using five key indicators: F-measure, AUC, Percent of Perfect Cleans (PPC), Percent of Non-Perfect Cleans (PNPC), and False Omission Rate (FOR). Although F-measure and AUC provide an overall assessment of predictive performance, they can not fully capture the effectiveness in practical SDP scenarios. To address this gap, we introduce PPC, PNPC, and FOR.
For the second goal, we compare the F-measure and AUC of Joint\_CNN with existing explainable and non-explainable SDP models. PPC, PNPC and FOR are defined as follows:
\begin{equation}
    PPC = \frac{\left| TN \right|}{\left| D \right|} 
    \label{eq_PPC}
\end{equation}
\begin{equation}
    PNPC = \frac{\left| TP \right| + \left| FP \right| + \left| FN \right|}{\left| D \right|}
    \label{eq_PNPC}
\end{equation}
\begin{equation}
    FOR = \frac{\left| FN \right|}{\left| TN \right| + \left| FN \right|}
    \label{eq_FOR}
\end{equation}
where $\left| TP \right|, \left| FP \right|, \left| FN \right|,\left| FN \right|$ represent the number of true positives, false positives, true negative and false negatives.

The F-measure assesses the balance between precision and recall.
\begin{equation}
    F\text{-measure} = \frac{2 \cdot \text{Precision} \cdot \text{Recall}}{\text{Precision} + \text{Recall}}
    \label{eq_Fmeasure}
\end{equation}

The AUC is used to evaluate the performance of binary classification models by measuring the area under its Receiver Operating Characteristic (ROC) curve.
\textbf{RQ3:} what's the performance of our method in terms of global interpretability?

The RQ3 involves three objectives: (1) evaluating the global interpretability of our approach, (2) assessing its computational efficiency, and (3) demonstrating its reliability. 
For the first objective, we aim to demonstrate that our approach provides decision tree-like interpretability. The decision tree is a typical global interpretation method that represents the overall decision logic through hierarchical ``IF-Then'' rules \cite{60}. 
Furthermore, we adopt the same quantitative method as \cite{covert2020understanding} to evaluate global interpretability by measuring the contributions of metric subsets, identified by the interpreter, to the model's predictive performance. Specifically, we use Joint\_SDT and SP-LIME to generate six metric subsets for each dataset and use them to retrain Joint\_CNN independently. The predictive performance of Joint\_CNN with different metric subsets is measured by AUC. Higher AUC values indicate that the interpreter provides better global interpretability.
For the second objective, we assess the computational efficiency of our method and SP-Lime by comparing the training times. Since excessive training time is a key limitation of existing global interpretation methods \cite{61}.
For the third objective, we use three indicators, CD, FI and AI to quantify the reliability of our approach.

\textbf{RQ4:} Do the metrics selected by Joint\_SDT have a significant impact on the performance of Joint\_CNN?

This study evaluates the consistency between the metrics selected by Joint\_SDT and those impact the  performance of Joint\_CNN most. To address this, we conduct ablation experiments in which each metric identified by Join\_SDT is individually removed, and the resulting degradation in Joint\_CNN’s performance (e.g., declines in Matthews Correlation Coefficient (MCC), F-measure, and AUC) is measured.
MCC is commonly used to assess the quality of imbalanced binary classification, where the value ranges from -1 to 1. -1 indicates complete disagreement between predictions and ground truth, 1 represents perfect agreement.
A strong correlation between Joint\_SDT’s prioritized metrics and the magnitude of performance deterioration would further confirm that our joint-learning framework make positive contributions to both predictor and interpreter.

\begin{equation}
    MCC = \frac{TP \cdot TN - FP \cdot FN}{\sqrt{(TP + FP)(TP + FN)(TN + FP)(TN + FN)}}
    \label{eq_Mcc}
\end{equation}

\begin{figure*}[!t]
    \centering
    \subfloat[The first interpretation]{
        \includegraphics[width=0.35\textwidth, height=0.12\textheight]{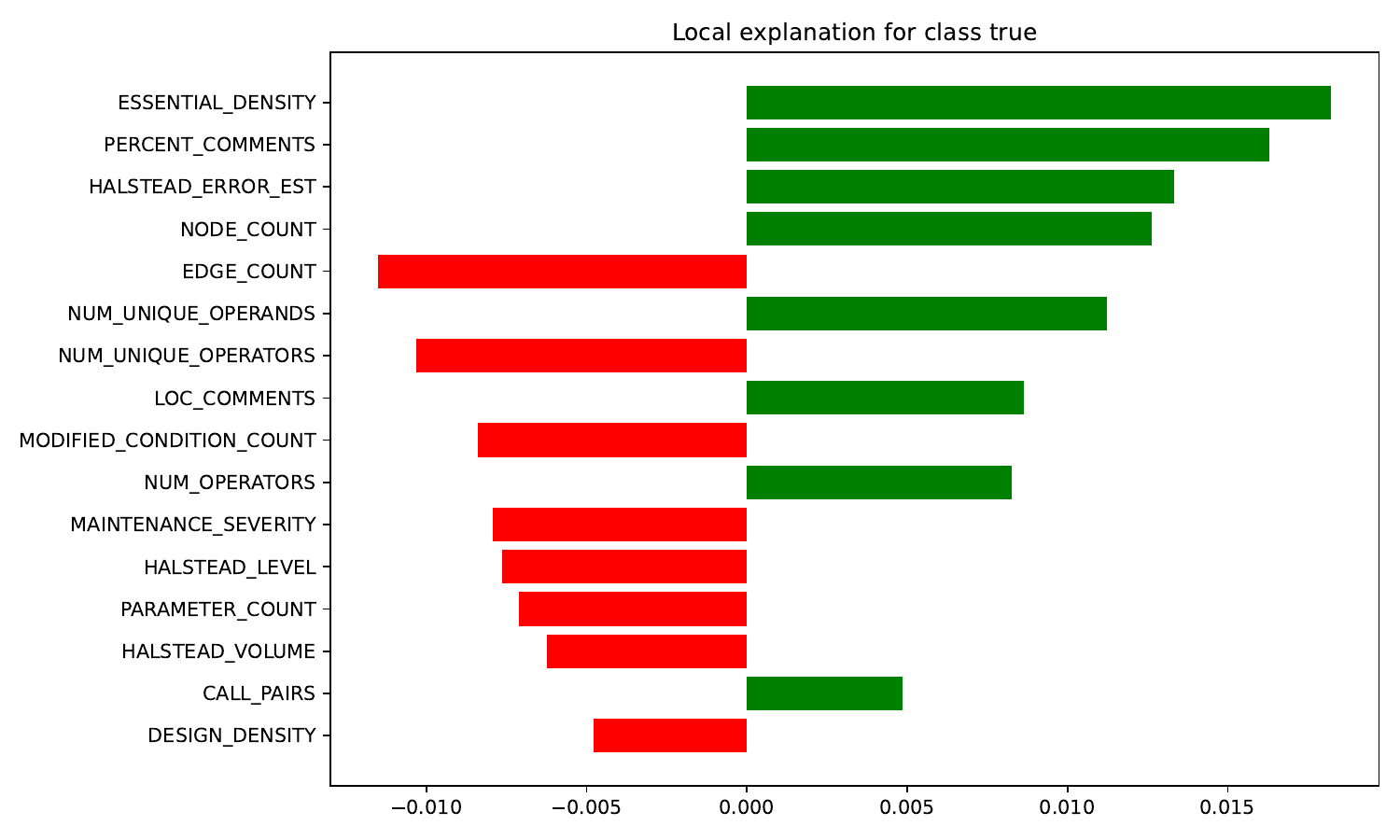}
        \label{fig4:Interpretation1}
    }
    \hspace{0.05\textwidth} 
    \subfloat[The second interpretation]{
        \includegraphics[width=0.35\textwidth, height=0.12\textheight]{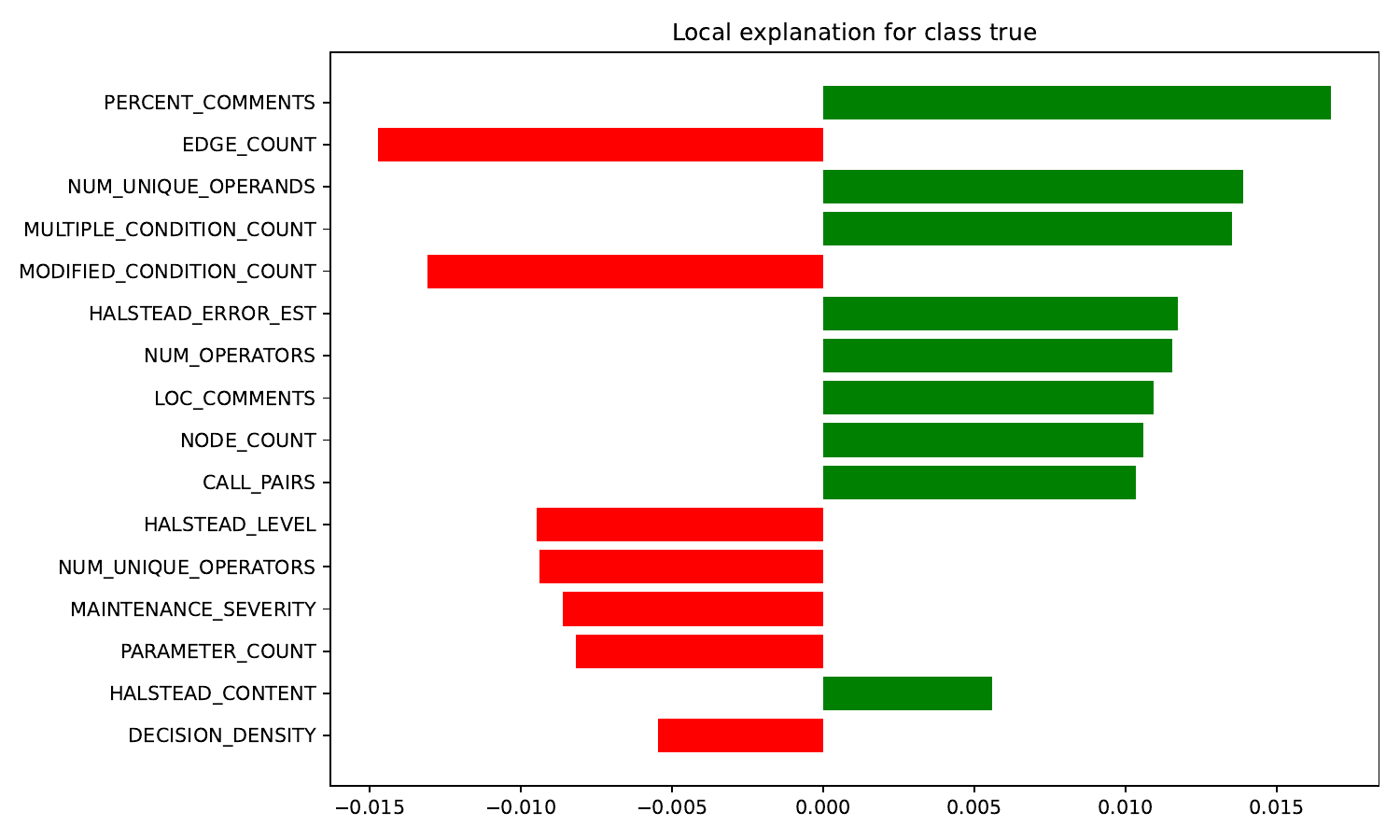}
        \label{fig4:Interpretation2}
    }

    \subfloat[The third interpretation]{
        \includegraphics[width=0.35\textwidth, height=0.12\textheight]{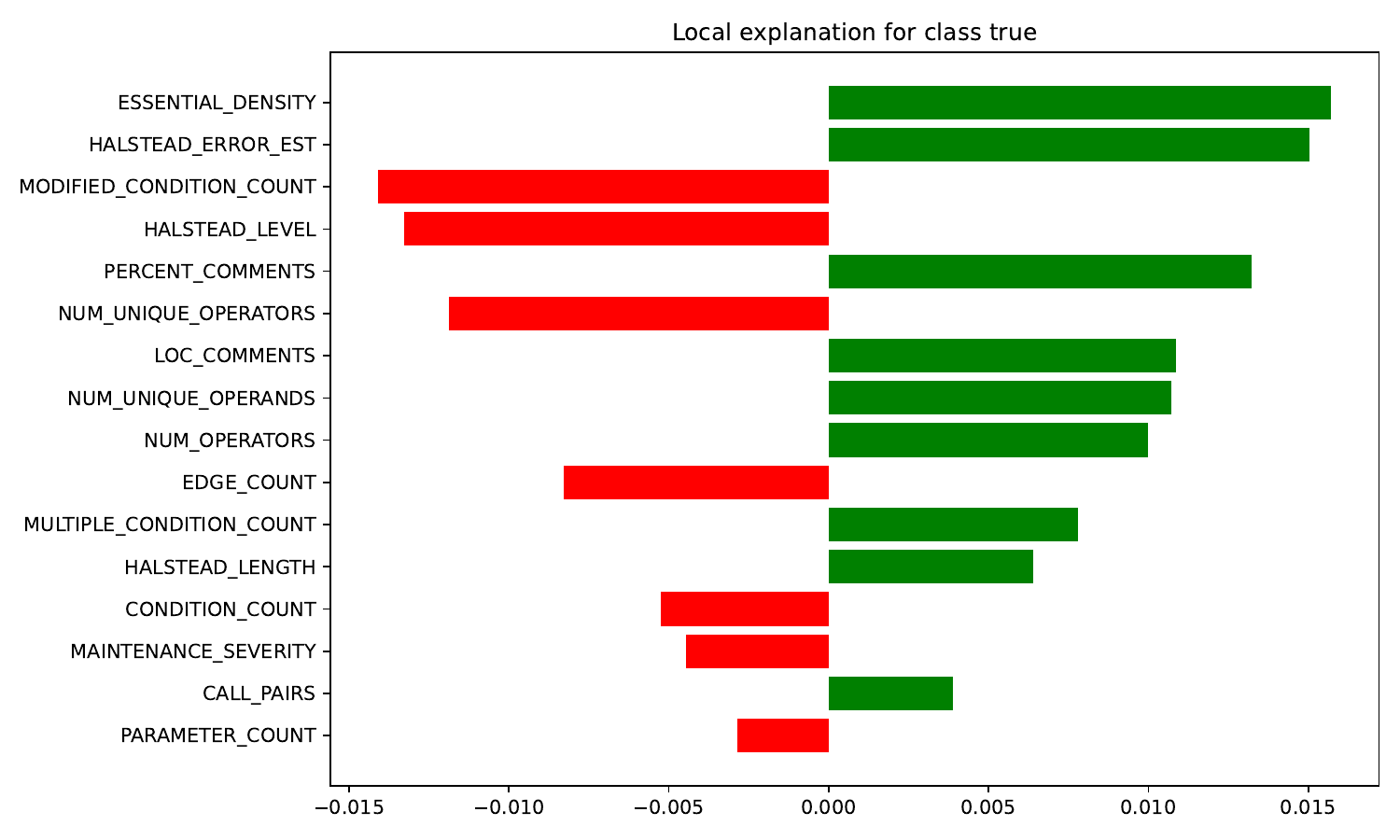}
        \label{fig4:Interpretation3}
    }
    \hspace{0.05\textwidth} 
    \subfloat[The fourth interpretation]{
        \includegraphics[width=0.35\textwidth, height=0.12\textheight]{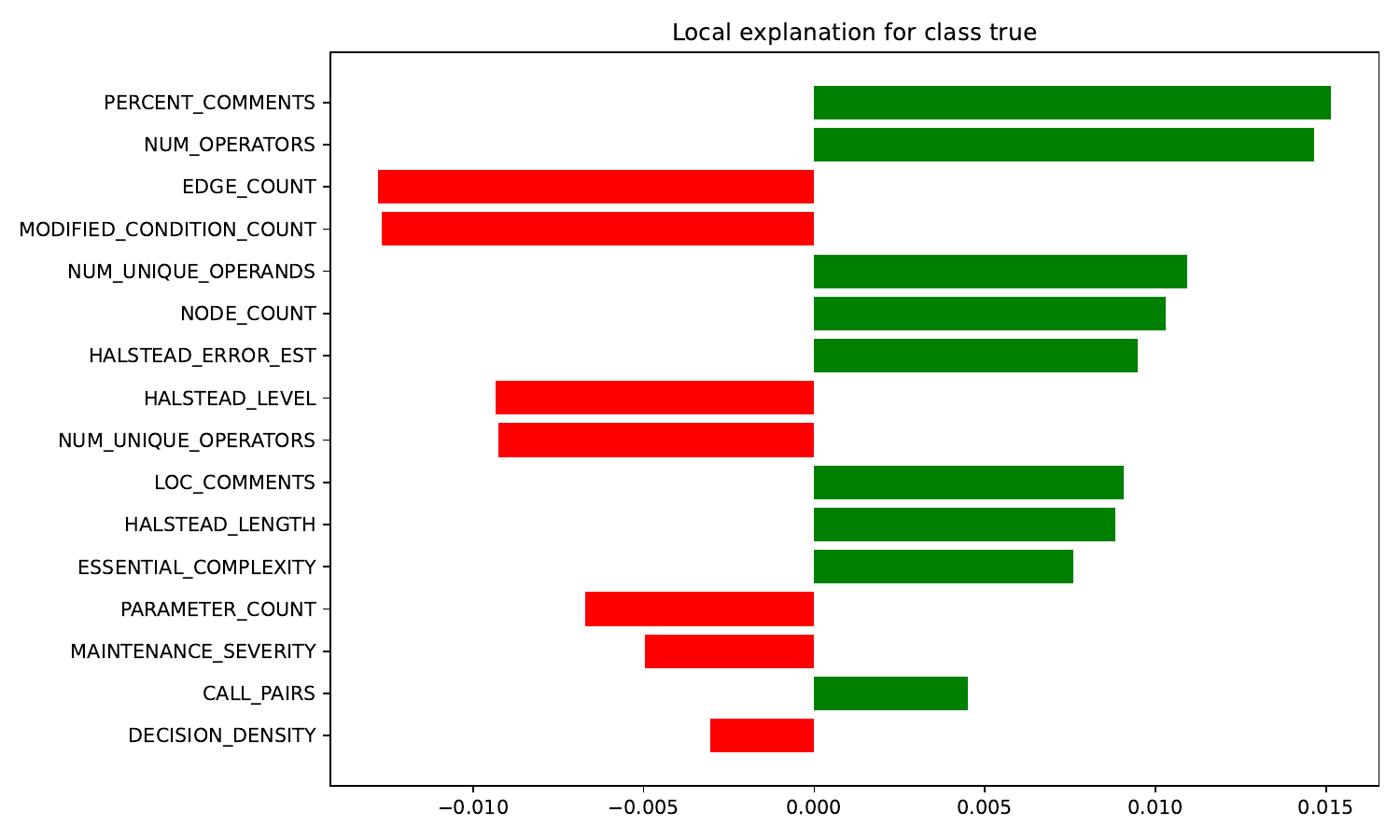}
        \label{fig4:Interpretation4}
    }
 
    \caption{Multiple interpretations of LIME for a certain instance in the PC2 dataset.}
    \label{fig4}
\end{figure*}

\subsection{Dataset}
Twenty datasets used in the experiment are collected from the AEEEM \footnote{ https://zenodo.org/record/3362613\#.YmuGoNpByUk}, NASA \footnote{http://promise.site.uottawa.ca/SERepository/datasets-page.html} and PROMISE \footnote{http://promise.site.uottawa.ca/SERepository} repositories. Numerous studies have 	demonstrated the efficacy of these datasets \cite{51, 52, 53, 54}. Further details about each dataset are provided in Table \ref{tab2}. We observe that these datasets exhibit significant class imbalance, with the highest imbalance rate reaching 0.41\% from the PC2 dataset.

\begin{table}[!t]
\centering
\caption{Details of the datasets used in the experiment}
\begin{tabular}{lllll}
\toprule
Group & Dataset & Instances & Defects    &\%Defects \\
\midrule
AEEEM       & JDT           & 997       & 206        &    20.66\% \\
            & LC            & 691       & 64         &    9.26\% \\
            & ML            & 1862      & 245        &    13.16\% \\
            & PDE           & 1492      & 209        &    14.01\% \\
\midrule
NASA        & kc1           & 2109      & 326        &    15.46\% \\
            & mc1           & 9466      & 68         &    0.72\% \\
            & jm1           & 7782      & 1672       &    21.49\% \\
            & PC1           & 549       & 77         &    14.03\% \\
            & PC2           & 5589      & 23         &    0.41\% \\
            & PC3           & 1563      & 160        &    10.24\% \\
            & PC4           & 1458      & 178        &    12.21\% \\
\midrule 
PROMISE     & ant-1.7       & 745       & 166        &    22.28\% \\
            & camel-1.6     & 965       & 188        &    19.48\% \\
            & ivy-1.2       & 352       & 40         &    11.36\% \\
            & jedit-4.1     & 312       & 79         &    25.32\% \\
            & log4j-1.0     & 135       & 34         &    25.19\% \\
            & lucene-2.4    & 340       & 203        &    59.71\% \\
            & poi-3.0       & 442       & 281        &    63.57\% \\
            & synapse1.2    & 256       & 86         &    33.59\% \\
            & xerces-1.3    & 453       & 69         &    15.23\% \\

\bottomrule
\end{tabular}
\label{tab2}
\end{table}

\subsection{Setting}
We config our model as follows: the learning rate is initialized as 1e-06 and the batch size is set to 16.  SGD optimizer is employed with a momentum of 0.9. The depth of SDT is set to 4. The penalty strength of SDT is set to 1e+1 and the penalty decay rate is set to 0.25. The hyperparameters $\alpha$ and $\beta$ in Eq. (2) are set to 1.4 and 0.6, and $\gamma$, $\lambda$ in Eq. (2) are set to 0.8 and 0.5. The temperature hyperparameter \textit{T} in Eq. (6) is set to 100. The window size for the Exponential Moving Average (EMA) is set to 1000.  Training stops when the model reaches the maximum number of iterations (40) or when early stopping criterias are met (no validation loss improvement for 10 consecutive epochs). \textcolor{blue}{Based on preliminary experimental results, the maximum number of iterations is set to 40, beyond which model performance saturates with minimal further improvement.}
The dataset was split into 7:2:1 for training, testing, and validation. The experiments are implemented by TensorFlow 2.3.0 with the NVIDIA GeForce RTX 3090 GPU. To ensure fairness, we applied this configuration to the experiments on all datasets.

\subsubsection{Baseline}
For RQ1, we utilize LIME \cite{lime} and BreakDown \cite{34} as baselines for the following reasons: (1) These methods are extensively used in the SDP research; (2) LIME, BreakDown, and our approach belong to feature-oriented interpretation techniques that aim to quantify the correlations between input features and predictive outputs; (3) The theoretical foundations of LIME and BreakDown are different, thereby enriching the diversity of comparisons. For RQ2, we use SVM, Random Forest (RF), Deep Belief Network (DBN), and AutoSpearman \cite{18} as baselines. These models are selected based on the following reasons: (1) SVM and RF are widely adopted non-explainable SDP models, while DBN represents DL-based approach in SDP. (3) AutoSpearman, in contrast, is a well-recognized explainable SDP model. Including it as a baseline ensures a fair comparison. For RQ3, we use SP-LIME \cite{60} as the baseline for two key reasons: (1) SP-LIME and our method share the same technical foundation. Both of them are perturbation-based global interpretable methods. Moreover, SP-LIME is the representative method in this category \cite{60}. (2) SP-LIME is the global version of Lime (one of the baselines of RQ1). Selecting it as the baseline is helpful to maintain the consistency across the experiments.

\subsubsection{Data preprocessing}
It involves normalization and data sampling in this section. Specifically, normalization is defined as follows:
\begin{equation}
    \widetilde{x}=\frac{x-\min{\left(D\right)}}{\max{\left(D\right)}-\min{\left(D\right)}}
\end{equation}
where $x$ is the sample of a dataset, and ${\rm min}(\cdot), {\rm max}(\cdot)$ are the minimum and maximum operations applied column-wise across the dataset. Normalization ensures that all features are on the same scale, and it helps the algorithm converge faster and more smoothly. To address the class imbalance issue, we adopt two sampling techniques: SMOTE (Synthetic Minority Oversampling Technique) \cite{55} and RUS (Random Undersampling) \cite{RUS}. 

\subsubsection{Statistical significance test}
To verify the significance of the improvements, we introduce $Cohen's\ d$ in this section, which takes any value between 0 and infinity. When $Cohen's\ d$ takes values in the ranges $[0,0.2), [0.2,0.8), and\ [0.8, +\infty)$, it indicates that the improvement is small, medium, and large \footnote{ https://en.wikipedia.org/wiki/Effect\_size}.
The definition of $Cohen's\ d$ is as follows:
\begin{equation}
    Cohen's\ d = \frac{M_1 - M_2}{\sqrt{(SD_1^2 + SD_2^2)/2}}
\end{equation}
where $M_1$ and $M_2$ represent the means of the two group data, and $SD_1$ and $SD_2$ represent their standard deviations.

\begin{table}[!t]
\centering
\caption{The \textit{CD-10\%}s of LIME, BreakDown and proposed method}
\begin{tabular}{llll}
\toprule
        &LIME & BreakDown & Joint\_SDT \\
\midrule
Average CD-10\%  &65\% &74.5\% &100\% \\
\midrule
Improvement &53.85\%\textuparrow &34.23\%\textuparrow &- \\
\bottomrule
\end{tabular}
\label{CDs}
\end{table}

\begin{table}[!t] 
\centering
\caption{The AI and FI of Joint\_SDT and LIME}
\begin{tabular}{lllll}
\toprule
& \multicolumn{2}{l}{$AI$} & \multicolumn{2}{l}{$FI$} \\
\cmidrule(ll){2-3} \cmidrule(ll){4-5}
& LIME & Joint\_SDT & LIME  & Joint\_SDT \\
\midrule
JDT       & 0.8200  & 0.8200           & 0.8100   & \textbf{0.9300}  \\
LC        & 0.8143  & \textbf{0.8429}  & 0.8000   & \textbf{0.8714}  \\
ML        & 0.6684  & \textbf{0.9037}  & 0.6791   & \textbf{0.9786}  \\
PDE       & 0.7200  & \textbf{0.7533}  & 0.6533   & \textbf{0.7867}  \\
jm1       & 0.6239  & \textbf{0.8010}  & 0.8793   & \textbf{0.9910}  \\
kc1       & 0.6540  & \textbf{0.8531}  & 0.5355   & \textbf{0.9431}  \\ 
mc1       & 0.8754  & \textbf{0.9916}  & 0.9155   & \textbf{0.9979}  \\
PC1       & 0.8727  & 0.8727           & 0.7091   & \textbf{0.9273}  \\
PC2       & 0.9821  & \textbf{0.9982}  & 0.9714   & \textbf{0.9803}  \\
PC3       & 0.7834  & \textbf{0.9172}  & 0.8471   & \textbf{0.9682}  \\
PC4       & 0.8151  & \textbf{0.9178}  & 0.9041   & \textbf{0.9932}  \\
ant-1.7    & 0.6667  & \textbf{0.7467}  & 0.6667  & \textbf{0.9200}  \\
camel1.6   & 0.7010 & 0.5361          & 0.7010  & \textbf{0.8144}  \\
ivy-1.2     & 0.8056 & \textbf{0.8333}  & 0.8056 & \textbf{0.9167}  \\
jedit-4.1   & 0.5312  & \textbf{0.7500}  & 0.5312  & \textbf{0.5938}  \\
log4j-1.0   & 0.7857  & 0.6429           & 0.7857  & 0.5714  \\
lucene-2.4  & 0.5000  & \textbf{0.7059}  & 0.5000  & \textbf{0.7059}  \\
poi-3.0     & 0.4222  & \textbf{0.7778}  & 0.4222  & \textbf{0.8222}  \\
synapse-1.2 & 0.5769  & \textbf{0.6538}  & 0.5769  & 0.5769  \\
xerces-1.3  & 0.6522  & \textbf{0.7174}  & 0.6522  & \textbf{0.8478}  \\
\midrule
Average     & 0.7135 & \textbf{0.8018} & 0.7173  & \textbf{0.8568}  \\
\midrule
Improvement & 12.38\%\textuparrow & - & 19.45\%\textuparrow & - \\
\midrule
$Cohen's\ d$& 0.6792 & -    & 0.9451  & -      \\
\bottomrule
\end{tabular}
\label{tab:lime_JointSDT}
\end{table}

\begin{table*}[!t]
\centering
\caption{Prediction Performance of Joint\_CNN and Base\_CNN in Terms of F-measure, AUC, PPC, PNPC and FOR}
\resizebox{\textwidth}{!}{%
\begin{tabular}{lllllllllll}
\toprule
& \multicolumn{2}{l}{F-measure} & \multicolumn{2}{l}{AUC} & \multicolumn{2}{l}{PPC}& \multicolumn{2}{l}{PNPC}& \multicolumn{2}{l}{FOR}\\
\cmidrule(ll){2-3} \cmidrule(ll){4-5}\cmidrule(ll){6-7}\cmidrule(ll){8-9}\cmidrule(ll){9-11}
& Base\_CNN &Joint\_CNN  &Base\_CNN &Joint\_CNN &Base\_CNN &Joint\_CNN &Base\_CNN &Joint\_CNN &Base\_CNN &Joint\_CNN \\
\midrule
JDT & 0.6802 & \textbf{0.8544} & 0.7297 & \textbf{0.8360} & 0.6000 &\textbf{0.6900}     &0.4000    &\textbf{0.3100} &0.2683    &\textbf{0.1585} \\
LC  & 0.5686 & \textbf{0.8285} & 0.6923 & 0.6862 & 0.7714  & \textbf{0.7857}            &0.2286    &\textbf{0.2143} &0.1692    &\textbf{0.1538} \\
ML  & 0.5686 & \textbf{0.7123} & 0.6897 & 0.6418 & 0.6203 & \textbf{0.6952}             &0.3797    &\textbf{0.3048} &0.3216    &\textbf{0.2398} \\
PDE & 0.7475 & 0.7111          & 0.7039 & 0.7032  & 0.7000  & \textbf{0.7867}           &0.3000    &\textbf{0.2133} &0.2105    &\textbf{0.1128} \\
jm1 & 0.5755 & \textbf{0.8621} & 0.6240 & \textbf{0.6280} &0.5417  & \textbf{0.8036 }   &0.4583    &\textbf{0.1964} &0.1335    &\textbf{0.0012} \\
kc1 & 0.6443 & \textbf{0.7303} & 0.7545 & \textbf{0.8087} & 0.5024  & \textbf{0.5308}   &0.4976    &\textbf{0.4692} &0.0783    &\textbf{0.0345} \\
mc1 & 0.7352 & \textbf{0.8621} & 0.8082 & 0.6280 & 0.8786  & 0.7818                     &0.1214    &0.2182          &0.0012    &0.0652 \\
PC1 & 0.7475 & \textbf{0.8621} & 0.7039 & 0.6280 &0.6909  &\textbf{0.8182 }             &0.3091    &\textbf{0.1818} &0.0500    &0.0816 \\
PC2 & 0.4996 & \textbf{0.8621} & 0.5000 & \textbf{0.6280} & 0.9911 & 0.9875             &0.0089    &0.0125          &0.0018    &0.0018 \\
PC3 & 0.5588 & \textbf{0.7957} & 0.7634 & \textbf{0.8324} & 0.7707 & 0.6752             &0.2293    &0.3248          &0.0397    &\textbf{0.0093} \\
PC4 & 0.7021 & \textbf{0.7472} & 0.8169 & 0.7843 & 0.8082 & 0.6644                      &0.1918    &0.3356          &0.0167    &0.0300 \\
ant-1.7      & 0.7991 & 0.7907 & 0.8276 & 0.8022 & 0.7733 & 0.5733                      &0.2267    &0.4267          &0.2267    &\textbf{0.0444}\\
camel-1.6    & 0.6932 & \textbf{0.7337} & 0.5992 & \textbf{0.6687} &0.5567 &\textbf{0.5876}  &0.4433    &\textbf{0.4124} &0.1692    &0.1739 \\
ivy-1.2      & 0.8273 & \textbf{0.8347} & 0.6016 & \textbf{0.8594} &0.7778 &0.7500           &0.2222    &0.2500          &0.0667    &0.1000 \\
jedit-4.1    & 0.5862 & \textbf{0.6946} & 0.5429 & \textbf{0.8125} & 0.5312 &\textbf{0.5938} &0.4688    &\textbf{0.4062} &0.0556    &\textbf{0.0500} \\
log4j-1.0    & 0.7755 & 0.6520          & 0.8500 & 0.8500 &0.5714 &0.3571                    &0.4286    &0.6429          &0.2727    &\textbf{0.0235}\\
lucene-2.4   & 0.5623 & \textbf{0.6947} & 0.5482 & \textbf{0.7643} &0.2941 &\textbf{0.3529}  &0.7059    &\textbf{0.6471} &0.4737    &\textbf{0.3684} \\
poi-3.0      & 0.7156 & 0.6951          & 0.7263 & \textbf{0.7414} &0.4000 &\textbf{0.5333}  &0.6000    &\textbf{0.4667} &0.3793    &\textbf{0.1724} \\
synapse1.2   & 0.6616 & 0.6486          & 0.8235 & 0.7516 &0.3077          &\textbf{0.6154}  &0.6923    &\textbf{0.3846} &0.2000    &\textbf{0.1111} \\
xerces-1.3   & 0.7921 & \textbf{0.8407} & 0.8498 & 0.7875 &0.6304          &\textbf{0.6957}  &0.3696    &\textbf{0.3043} &0.1212    &\textbf{0.1111} \\
\midrule
Average   & 0.6720  & \textbf{0.7706}  & 0.7078   & \textbf{0.7421} &0.6359  &\textbf{0.6639 } & 0.3641  & \textbf{0.3361} & 0.1628  &\textbf{0.1022 }\\
\midrule
Improvement & 14.67\%\textuparrow & - & 4.85\%\textuparrow & - &4.4\%\textuparrow&- &8.33\% \textuparrow &-&59.30\%\textuparrow &-\\
\midrule
$Cohen's\ d$  & 1.1321 &- &0.3553 & - &0.0763  &-  &0.1665 &-  &0.5338 &- \\
\bottomrule
\end{tabular}
}
\label{tab:joint_Base_CNN}
\end{table*}

\begin{table*}[t]
\centering
\caption{The F-measure and AUC performance of different models on all datasets}
\begin{tabular}{lllllllllll}
\toprule
& \multicolumn{5}{l}{F-measure} & \multicolumn{5}{l}{AUC} \\
\cmidrule(lllll){2-6} \cmidrule(lllll){7-11}
Dataset    & SVM    & RF     & DBN    & AutoSpearman & Joint\_CNN   & SVM    & RF     & DBN    & AutoSpearman & Joint\_CNN \\
\midrule
JDT        & 0.6239 & 0.6989 & 0.7162 & 0.7824    & \textbf{0.8544}      & 0.7554 & 0.7419 & 0.7852 & 0.8598           & 0.8360    \\
LC         & 0.3467 & 0.7177 & 0.3467 & 0.7871    & \textbf{0.8285}      & 0.5846 & 0.8462 & 0.5846 & 0.5046           & 0.6862    \\
ML         & 0.4919 & 0.5958 & 0.3219 & 0.8054    & 0.7123               & 0.6732 & 0.7522 & 0.4596 & 0.7339           & 0.6418    \\
PDE        & 0.5565 & 0.6209 & 0.4874 & 0.7837    & 0.7111               & 0.6875 & 0.6601 & 0.6130 & 0.6957           & \textbf{0.7032}    \\
jm1        & 0.4476 & 0.5849 & 0.4476 & 0.7580    & \textbf{0.8621}      & 0.5030 & 0.5762 & 0.5030 & 0.6269           & \textbf{0.6280}    \\
kc1        & 0.5695 & 0.5371 & 0.6918 & 0.5814    & \textbf{0.7303}      & 0.6992 & 0.6141 & 0.7724 & 0.8111           & 0.8087    \\
mc1        & 0.4754 & 0.5513 & 0.4742 & 0.5814    & \textbf{0.8621}      & 0.8914 & 0.5577 & 0.8326 & 0.9373           & 0.6280    \\
PC1        & 0.3870 & 0.7556 & 0.3939 & 0.8276    & \textbf{0.8621}      & 0.6562 & 0.8557 & 0.6057 & 0.6815           & 0.6280    \\
PC2        & 0.4736 & 0.4996 & 0.4213 & 0.9819    & 0.8621               & 0.4507 & 0.5000 & 0.3647 & 0.4194           & \textbf{0.6280}    \\
PC3        & 0.5580 & 0.5745 & 0.5224 & 0.8345    & 0.7957               & 0.7914 & 0.5772 & 0.7356 & 0.8336           & 0.8324    \\
PC4        & 0.5700 & 0.6694 & 0.3535 & 0.8617    & 0.7472               & 0.7380 & 0.6154 & 0.5651 & 0.7814           & \textbf{0.7843}    \\
ant-1.7    & 0.7900 & 0.7121 & 0.5446 & 0.7760    & \textbf{0.7907}      & 0.8409 & 0.7477 & 0.6591 & 0.8266           & 0.8022    \\
camel-1.6  & 0.4920 & 0.5367 & 0.6314 & 0.6710    & \textbf{0.7337}      & 0.6199 & 0.5621 & 0.6923 & 0.6356           & 0.6687    \\
ivy-1.2    & 0.6244 & 0.7317 & 0.5000 & 0.8721    & 0.8347               & 0.6719 & 0.8125 & 0.7500 & 0.6250           & \textbf{0.8594}    \\
jedit-4.1  & 0.3914 & 0.3273 & 0.3431 & 0.7585    & 0.6946               & 0.4657 & 0.3743 & 0.4771 & 0.8698           & 0.8125    \\
log4j-1.0  & 0.4286 & 0.3538 & 0.4286 & 0.6533    & 0.6520               & 0.6364 & 0.5909 & 0.6364 & 0.7750           & \textbf{0.8500}    \\
lucene-2.4 & 0.7153 & 0.3585 & 0.5983 & 0.7059    & 0.6947               & 0.7140 & 0.5000 & 0.6333 & 0.8250           & 0.7643    \\
poi-3.0    & 0.6584 & 0.6786 & 0.5139 & 0.7819    & 0.6951               & 0.6463 & 0.6624 & 0.5392 & 0.8082           & 0.7414    \\
synapse-1.2& 0.5439 & 0.6067 & 0.3210 & 0.5247    & \textbf{0.6486}      & 0.5437 & 0.6062 & 0.4125 & 0.6732           & \textbf{0.7516}    \\
xerces-1.3 & 0.6430 & 0.6865 & 0.4697 & 0.7611    & \textbf{0.8407}      & 0.7747 & 0.6758 & 0.6465 & 0.7766           & \textbf{0.7875}    \\

\midrule
Average   & 0.5394 & 0.5899 & 0.4764 & 0.7545 & \textbf{0.7706}  &0.6690  &0.6453  &0.6146 & 0.7350  & \textbf{0.7421} \\
\midrule
Improvement& 42.86\%\textuparrow & 30.63\%\textuparrow & 61.75\%\textuparrow & 2.13\%\textuparrow & -   &10.93\%\textuparrow &15.00\%\textuparrow &20.75\%\textuparrow & 0.97\%\textuparrow & - \\
\midrule
$Cohen's\ d$  &  2.3697 & 1.7196 & 2.9703  &0.1712 & - & 0.7337 & 0.9543 & 1.1906 &0.0657& - \\
\bottomrule
\end{tabular}
\label{tab:JointCNN_SVM_DBN_RF_AutoSpearman}
\end{table*}

\begin{table}[!t]
  \centering
  \caption{The CD, FI and AI values of Joint\_SDT and SP-Lime}
    \begin{tabular}{lcccc}
    \toprule
    Method    &CD   &FI  & AI    \\
    \midrule
    Joint\_SDT   & 1.00  & 0.8714   &  0.8429 \\
    SP-Lime   & 0.81    &  0.7933  &  0.8012 \\
    \bottomrule
    \end{tabular}%
  \label{tab77}
\end{table}%

\subsection{Experimental results}
The experimental results are presented as follows.
\subsubsection{The results of RQ1}
To evaluate the consistency of Joint\_SDT, LIME and BreakDown, we applied each of them to interpret the same instance ten times. As shown in Table \ref{CDs}, the average \textit{CD}-10\% scores of LIME, BreakDown and Joint\_SDT are 65\%, 74.5\%, and 100\%. This result suggests that both LIME and BreakDown exhibit poor consistency.
To further investigate the root cause of such poor consistency, we visualized the interpretations of LIME in Fig.~\ref{fig4}. We can see that LIME is unable to assign consistent weights to the same metric across multiple explanations, which we believe it is the root of poor consistency.
Moreover, we conducted a comparison between the interpretations generated by LIME and Joint\_SDT in terms of \textit{AI} and \textit{FI}. Since BreakDown does not provide an explicit surrogate function like LIME \cite{lime}, we are unable to calculate its \textit{AI} and \textit{FI}. As shown in Table \ref{tab:lime_JointSDT}, Joint\_SDT achieves an average improvement of 12.38\% in AI and 19.45\% in FI compared to LIME. Furthermore, the $Cohen's\ d$ results indicate that the significance of improvement in AI is medium, while in FI, it is large. This results further confirm that {Joint\_SDT} outperforms LIME in the consistency of interpretation. 

Although Joint\_SDT generally shows improvements in AI and FI, its performance is worse than LIME on certain datasets, such as Log4j. This is largely due to the limited amount of data, where sparse distributions enable LIME to maintain consistent interpretability by reducing sampling randomness. However, the neural network-like structure of Joint\_SDT requires a larger volume of data for stable and reliable feature importance estimation. As a result, Joint\_SDT struggles to achieve desirable performance on small datasets.

\subsubsection{The results of RQ2}

\textcolor{blue}{According to Table \ref{tab:joint_Base_CNN}, we observe that Joint\_CNN outperforms Base\_CNN in terms of F-measure, AUC, PPC, PNPC, and FOR on most datasets,  and the average improvements are 14.67\%,  4.85\%, 4.4\%, 8.33\% and 59.30\% respectively.} The corresponding $Cohen's\ d$ test results indicate that the effect sizes of the improvements in the five indicators are large, medium, small, small, and medium. From these results, we can draw the following conclusions: (1) The improvement of Joint\_CNN over Base\_CNN across all evaluation metrics indicates that the joint-learning framework makes a positive contribution in improving the performance of SDP models. (2) The improvements in PNPC and FOR suggest that Joint\_CNN is more reliable in distinguishing defective and non-defective instances, which is crucial for real-world SDP applications. (3) The constant improvements across all datasets suggests that the joint-learning framework generalize well and is not limited to specific dataset.


Furthermore, according to Table \ref{tab:JointCNN_SVM_DBN_RF_AutoSpearman}, joint\_CNN outperforms SVM, RF, DBN and AutoSpearman in both F-measure and AUC. Specifically, Joint\_CNN achieves average improvements of 42.86\%, 30.63\%, 61.75\% and 2.13\% in F-measure and 10.93\%, 15\%, 20.75\% and 0.97\% in AUC. The $Cohen's\ d$ test results indicate that the effect sizes of  of improvements in F-measure are large, large, large, and small, while those in AUC are medium, large, large, and small. According to these results, we can confirm that Joint\_CNN achieves better accuracy and demonstrates strong robustness across all dataset.

We believe there are two reasons related to the improvements: (1) We explicitly integrate the prediction error $L_{pred}(f,D)$ into the loss function of the joint-learning framework, guiding the predictor to capture discriminative information contained in the data. (2) The revised distillation loss  $L_{int}$ can be regarded as a penalty term of $L$, which effectively encourages the framework to learn more generalized features of the data, improving the model's robustness.

\subsubsection{The results of RQ3}
\begin{figure*}[ht]
    \centering
    \subfloat{
        \includegraphics[width=0.3\textwidth, height=0.15\textheight]{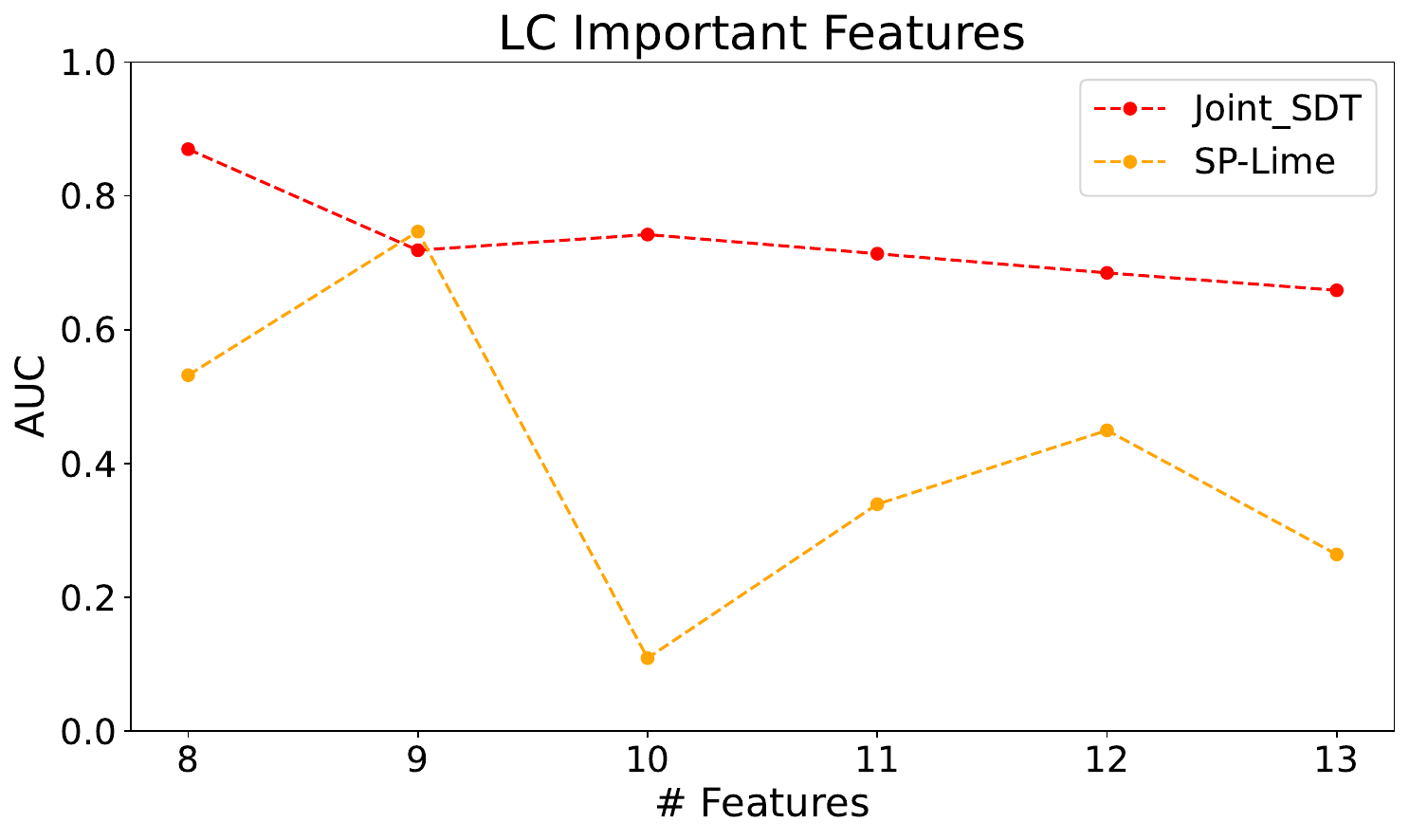}
    }
    \subfloat{
        \includegraphics[width=0.3\textwidth, height=0.15\textheight]{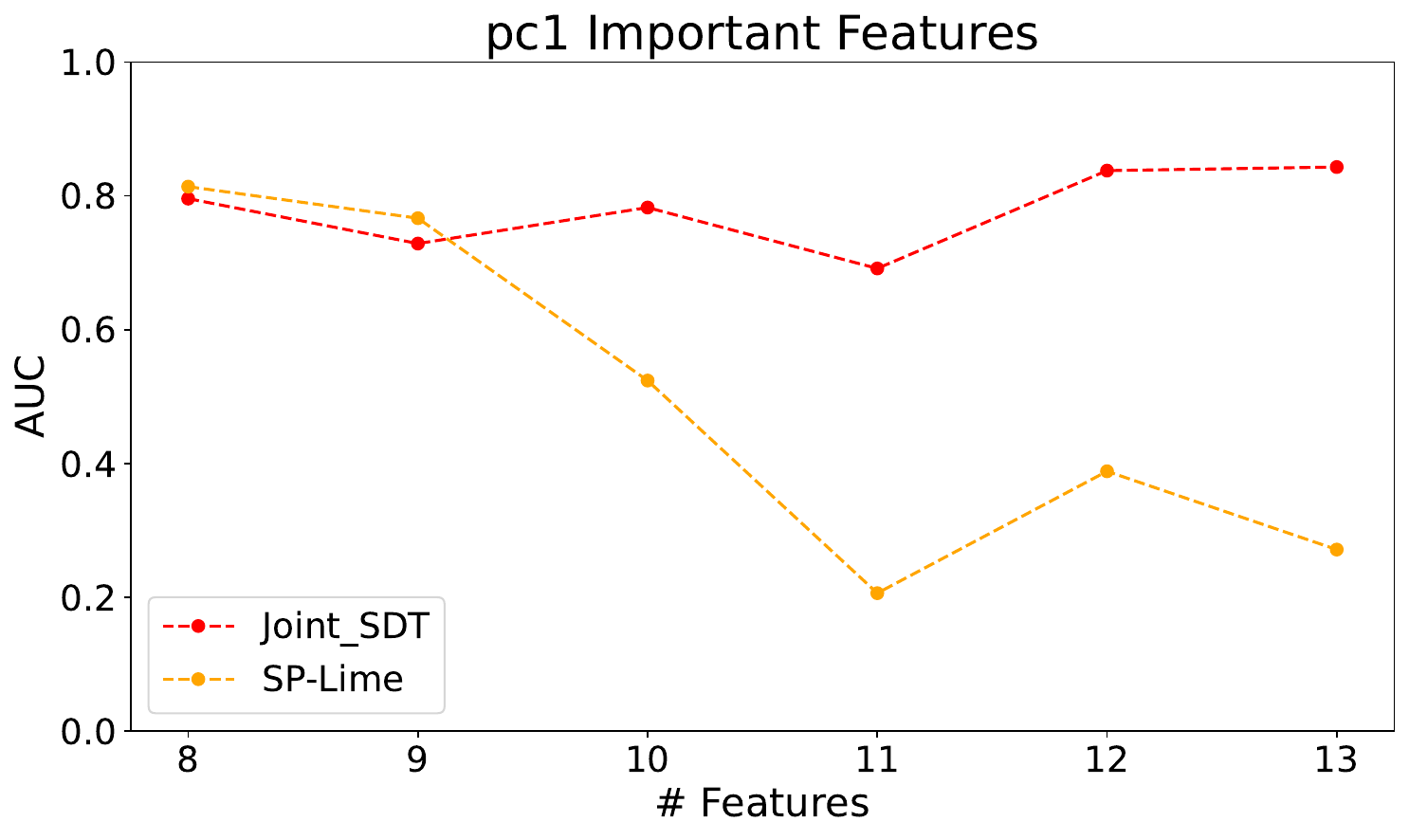}
    }
    \subfloat{
        \includegraphics[width=0.3\textwidth, height=0.15\textheight]{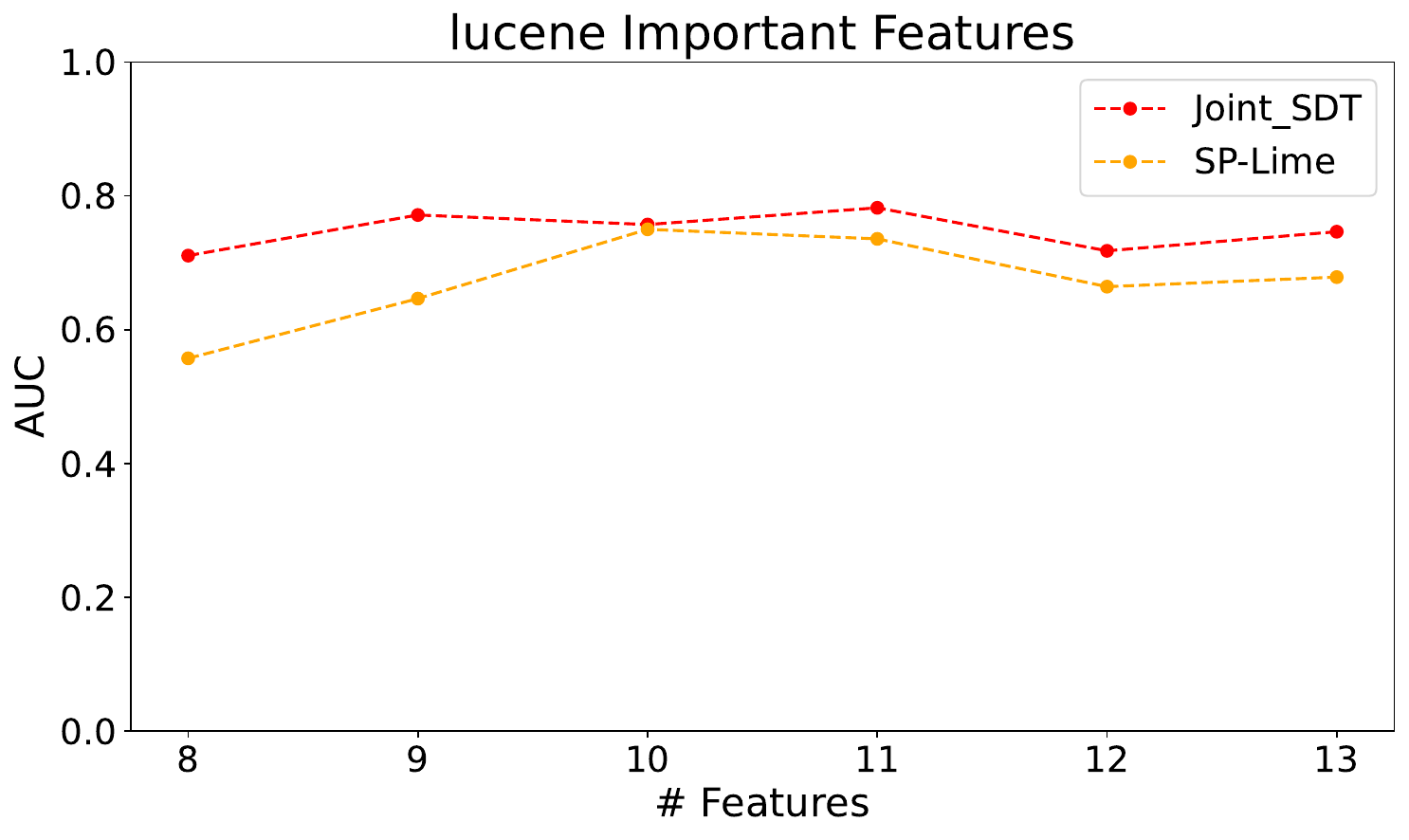}
    }
    \caption{Comparison of AUC values of Joint\_CNNs using metric subsets selected by Joint\_SDT and SP-LIME on LC, PC1, and Lucene datasets.}
    \label{sage}
\end{figure*}

To demonstrate the global interpretability of our approach, we use the PC1 dataset as an example. As presented in Fig.~\ref{PC2_tree}, the decision logic can be abstracted as the path from the ``\texttt{DECISION\_DENSITY}'' metric (denoted as the root) to the leaf node (denoted as 1 or 0). For example, one feasible decision path can be presented as \texttt{DECISION\_DENSITY -> MAINTENANCE\_SEVERITY -> PERCENT\_COMMENTS -> PERCENT\_COMMENTS -> 0}, which is similar to the interpretation format of a decision tree. 
Furthermore, following the framework of \cite{covert2020understanding}, we visualized the AUCs of Joint\_CNNs using different metric subsets identified by Joint\_SDT and SP-LIME in Fig.~\ref{sage}. The results demonstrate that our method achieves higher AUC values. This confirms that the metric subsets selected by Joint\_SDT are more effective for explaining the model’s global decision-making logic compared to those selected by SP-LIME.

We measure the computational cost by calculating the average training time. Specifically, we employ the Python $\mathrm{time}$ module to record the start and end times of the training process for both our approach and SP-Lime. To ensure statistical reliability, this procedure is repeated ten times for each method. The results show that the average computational costs for our approach and SP-Lime are 744.7 seconds and 2301.4 seconds, respectively. These findings indicate that our method achieves a significant reduction in computational cost compared to SP-Lime. The substantial difference can be attributed to the optimized design of our method, which abstracts the decision logic as the path of SDT from the root node to the leaf node. In contrast, SP-Lime requires extensive perturbations and evaluations across the feature space, leading to higher computational requirements.

Furthermore, we calculate the CD, FI and AI values for LC dataset to evaluate the reliability of our approach and SP-Lime. As presented in table \ref{tab77}, our method achieves a better performance in all three indicators. The reason for this result can be partially attributed to the random sampling involved in perturbation processes of SP-Lime \cite{60}.

\begin{figure*}[t]
    \centering
    \subfloat[F-measure]{
        \includegraphics[width=0.32\textwidth, height=0.2\textheight]{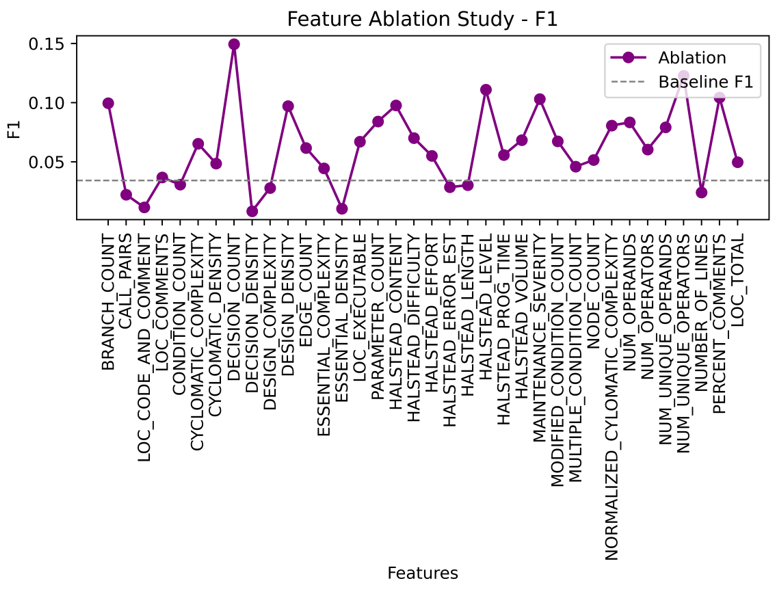}
        \label{fig:F1}
    } 
    \subfloat[AUC]{
        \includegraphics[width=0.32\textwidth, height=0.2\textheight]{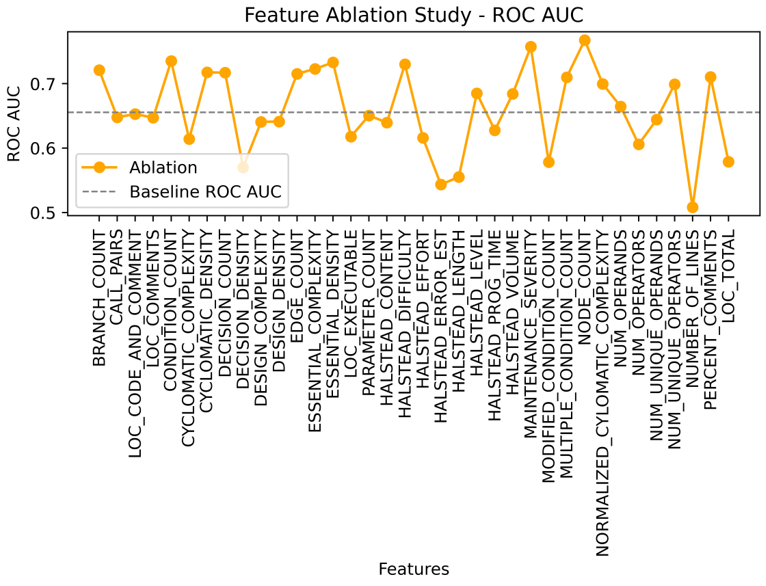}
        \label{fig:AUC}
    } 
    \subfloat[MCC]{
        \includegraphics[width=0.32\textwidth, height=0.2\textheight]{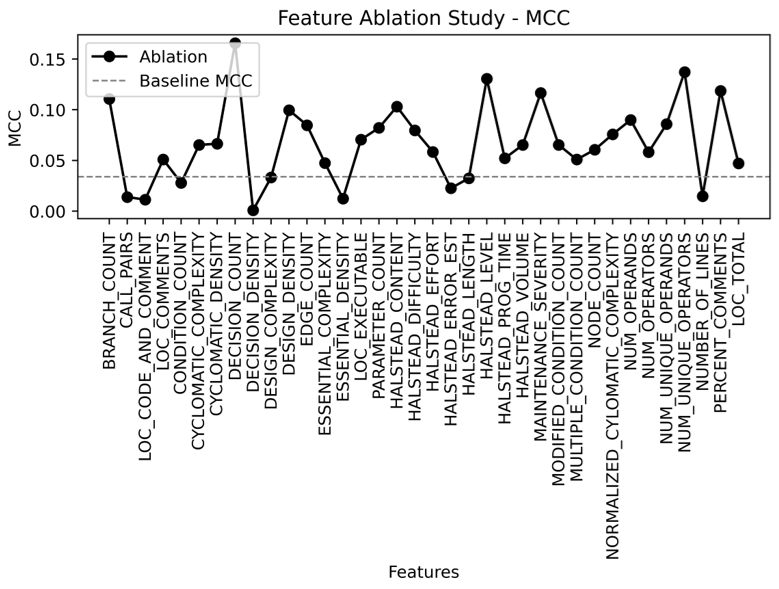}
        \label{fig:MCC}
    }
    \caption{Ablation study results across multiple indicators. The gray dashed line indicates the performance of DP-CNN before removing metrics.}
    \label{fig6}
\end{figure*}

\begin{table*}[t]
  \centering
  \caption{Rankings of Feature Importance}
    \begin{tabular}{lccccc}
    \toprule
    Dataset    &The Most Important Feature Identified by Joint\_SDT   &F-measure Rank  & AUC Rank    &MCC Rank  \\
    \midrule
    JDT   & WCHU\_numberOfMethodsInherited      & 4   & 3  & 5  \\
    LC    & WCHU\_cbo                           & 2   & 1  & 2  \\
    ML    & ck\_oo\_numberOfPublicMethods       & 3   & 2  & 5  \\
    PDE   & WCHU\_fanOut                        & 3   & 3  & 5  \\
    jm1   & HALSTEAD\_DIFFICULTY                & 3   & 2  & 1  \\
    kc1   & v                                   & 1   & 2  & 2  \\
    mc1   & LOC\_TOTAL                          & 1   & 1  & 1  \\
    PC1   & branchCount                  IV        & 2   & 1  & 1  \\
    PC2   & DECISION\_DENSITY                   & 1   & 4  & 1  \\
    PC3   & DESIGN\_DENSITY                     & 4   & 3  & 3  \\
    PC4   & LOC\_TOTAL                          & 2   & 2  & 1  \\

ant-1.7     & mfa     & 1       & 5       & 1  \\ 
camel-1.6   & lcom    & 4       & 3       & 4  \\ 
ivy-1.2     & cam     & 2       & 5       & 4  \\ 
jedit-4.1   & lcom3   & 4       & 5       & 4  \\ 
log4j-1.0   & max\_cc & 4       & 4       & 3  \\ 
lucene-2.4  & dam     & 1       & 1       & 1  \\
poi-3.0     & wmc     & 2       & 3       & 2  \\
synapse-1.2 & cbo     & 1       & 1       & 4  \\ 
xerces-1.3  & max\_cc & 2       & 5       & 2  \\  

\bottomrule
\end{tabular}%
\label{tab66}
\end{table*}%

\subsubsection{The results of RQ4}
According to the structure of Joint\_SDT, the most important metric is positioned at the root of the SDT.
For instance, in the PC2 dataset, "\texttt{DECISION\_DENSITY}" is identified as the most important metric. To validate this finding, we conduct an ablation experiment that removes each metric selected by Joint\_SDT individually and the corresponding changes are observed in the performance of Joint\_CNN. The ranking of performance degradation reflects the importance of each metric to the model. The ablation experiment results for PC2 are visualized in Fig. ~\ref{fig6}.
We can see that removing "\texttt{DECISION\_DENSITY}" results in the most significant performance decline across the three indicators: MCC, F-measure, and AUC. This confirms that "\texttt{DECISION\_DENSITY}" is the most important metric for PC2 dataset. This aligns with the interpretation of our method.

To enhance the comprehensiveness of the analysis, we extended the ablation experiment to all datasets. Table \ref{tab66} presents the most important metrics identified by Joint\_SDT across the twenty dataset. The results show that the root metrics identified by Joint\_SDT typically correspond to the most significant performance degradations in the ablation experiments.

\section{Threats to validity}
While the proposed model demonstrates improvements in both prediction accuracy and interpretability, certain limitations may affect its generalizability. The details are as follows.
\begin{itemize}
    \item  Choice of dataset: The experiments were conducted on the publicly available datasets. These datasets may not fully capture the diversity of software systems in real-world scenarios, potentially limiting the generalizability of the model.
    \item Hyperparameter setting: The model’s performance is related with the setting of hyperparameters for both the predictor and interpreter. Suboptimal setting may lead to less accurate predictions or interpretations.
    \item Application scenarios: In this paper, we focus on the within-project SDP. However, the performance of our model in the context of the cross-project setting has not been evaluated. 
\end{itemize}

\section{Conclusion and future works}\label{sec5}
In this paper, we explore the possibility of designing a defect predictor and its corresponding interpreter collaboratively. Unlike most existing approaches, which treat defect prediction and interpretation as separate tasks, we regard them as strongly correlated. We introduce a framework aiming at improving the reliability of interpretation and enhancing predictive accuracy simultaneously.
Based on the extensive empirical evaluations, we obtained the following key findings:
\begin{itemize}
    \item  Compared to existing interpretation methods such as LIME and BreakDown, the proposed framework significantly improves the reliability of interpretations through the incorporation of the KD principle.
    \item The joint-learning framework demonstrates superior accuracy on widely used datasets by explicitly incorporating interpretation results into the loss function.
\end{itemize}

Future research will explore the following directions: 
\begin{itemize}
    \item Using diverse real-world datasets to evaluate the robustness and generalization of our method.
    \item Integrating automatic hyperparameter optimization to reduce manual tuning efforts.
    \item Assess the model's performance in different defect prediction scenarios, such as cross-project defect prediction.
\end{itemize}

\bibliographystyle{IEEEtran}
\bibliography{IEEEabrv,references}

\begin{thebibliography}{10}
\providecommand{\url}[1]{#1}
\csname url@samestyle\endcsname
\providecommand{\newblock}{\relax}
\providecommand{\bibinfo}[2]{#2}
\providecommand{\BIBentrySTDinterwordspacing}{\spaceskip=0pt\relax}
\providecommand{\BIBentryALTinterwordstretchfactor}{4}
\providecommand{\BIBentryALTinterwordspacing}{\spaceskip=\fontdimen2\font plus
\BIBentryALTinterwordstretchfactor\fontdimen3\font minus \fontdimen4\font\relax}
\providecommand{\BIBforeignlanguage}[2]{{%
\expandafter\ifx\csname l@#1\endcsname\relax
\typeout{** WARNING: IEEEtran.bst: No hyphenation pattern has been}%
\typeout{** loaded for the language `#1'. Using the pattern for}%
\typeout{** the default language instead.}%
\else
\language=\csname l@#1\endcsname
\fi
#2}}
\providecommand{\BIBdecl}{\relax}
\BIBdecl

\bibitem{1}
W.~Zhang, Z.~Y. Ma, Q.~L. Lu, X.~B. Nie, and J.~Liu, ``Research on software defect prediction method based on machine learning,'' \emph{Applied Mechanics and Materials}, vol. 687, pp. 2182--2185, 2014.

\bibitem{2}
Q.~Wang, S.~Wu, and M.-S. Li, ``Software defect prediction,'' \emph{Journal of software}, vol.~19, no.~7, pp. 1565--1580, 2008.

\bibitem{55}
N.~V. Chawla, K.~W. Bowyer, L.~O. Hall, and W.~P. Kegelmeyer, ``Smote: synthetic minority over-sampling technique,'' \emph{Journal of artificial intelligence research}, vol.~16, pp. 321--357, 2002.

\bibitem{57}
C.~Lewis, Z.~Lin, C.~Sadowski, X.~Zhu, R.~Ou, and E.~J. Whitehead, ``Does bug prediction support human developers? findings from a google case study,'' in \emph{The 35th International Conference on Software Engineering}.\hskip 1em plus 0.5em minus 0.4em\relax IEEE, 2013, pp. 372--381.

\bibitem{3}
J.~Jiarpakdee, C.~K. Tantithamthavorn, and J.~Grundy, ``Practitioners' perceptions of the goals and visual explanations of defect prediction models,'' in \emph{The 18th International Conference on Mining Software Repositories}.\hskip 1em plus 0.5em minus 0.4em\relax IEEE, 2021, pp. 432--443.

\bibitem{4}
C.~K. Tantithamthavorn and J.~Jiarpakdee, ``Explainable ai for software engineering,'' in \emph{The 36th IEEE/ACM International Conference on Automated Software Engineering}.\hskip 1em plus 0.5em minus 0.4em\relax IEEE, 2021, pp. 1--2.

\bibitem{5}
H.~K. Dam, T.~Tran, and A.~Ghose, ``Explainable software analytics,'' in \emph{Proceedings of the 40th international conference on software engineering: New ideas and emerging results}, 2018, pp. 53--56.

\bibitem{murdoch2019definitions}
W.~J. Murdoch, C.~Singh, K.~Kumbier, R.~Abbasi-Asl, and B.~Yu, ``Definitions, methods, and applications in interpretable machine learning,'' \emph{Proceedings of the National Academy of Sciences}, vol. 116, no.~44, pp. 22\,071--22\,080, 2019.

\bibitem{7}
J.~Jiarpakdee, C.~Tantithamthavorn, and A.~E. Hassan, ``The impact of correlated metrics on defect models,'' \emph{arXiv preprint arXiv:1801.10271}, 2018.

\bibitem{YU2024}
J.~Yu, M.~Fu, A.~Ignatiev, C.~Tantithamthavorn, and P.~Stuckey, ``A formal explainer for just-in-time defect predictions,'' \emph{ACM Transactions on Software Engineering and Methodology}, vol.~33, no.~7, pp. 1--31, 2024.

\bibitem{58}
J.~Shin, R.~Aleithan, J.~Nam, J.~Wang, and S.~Wang, ``Explainable software defect prediction: Are we there yet?'' \emph{arXiv preprint arXiv:2111.10901}, 2021.

\bibitem{lime}
M.~T. Ribeiro, S.~Singh, and C.~Guestrin, ``{``Why should I trust you?'' Explaining the predictions of any classifier},'' in \emph{The 22nd ACM SIGKDD international conference on knowledge discovery and data mining}, 2016, pp. 1135--1144.

\bibitem{10}
S.~M. Lundberg and S.-I. Lee, ``A unified approach to interpreting model predictions,'' \emph{Advances in neural information processing systems}, vol.~30, 2017.

\bibitem{12}
J.~Bansiya and C.~G. Davis, ``A hierarchical model for object-oriented design quality assessment,'' \emph{IEEE Transactions on software engineering}, vol.~28, no.~1, pp. 4--17, 2002.

\bibitem{20}
I.-G. Czibula, G.~Czibula, Z.~Marian, and V.-S. Ionescu, ``A novel approach using fuzzy self-organizing maps for detecting software faults,'' \emph{Studies in Informatics and Control}, vol.~25, no.~2, pp. 207--216, 2016.

\bibitem{21}
D.~Kaur, A.~Kaur, S.~Gulati, and M.~Aggarwal, ``A clustering algorithm for software fault prediction,'' in \emph{International Conference on Computer and Communication Technology}.\hskip 1em plus 0.5em minus 0.4em\relax IEEE, 2010, pp. 603--607.

\bibitem{22}
Y.~Kamei, E.~Shihab, B.~Adams, A.~E. Hassan, A.~Mockus, A.~Sinha, and N.~Ubayashi, ``A large-scale empirical study of just-in-time quality assurance,'' \emph{IEEE Transactions on Software Engineering}, vol.~39, no.~6, pp. 757--773, 2012.

\bibitem{23}
X.~Yang, D.~Lo, X.~Xia, and J.~Sun, ``Tlel: A two-layer ensemble learning approach for just-in-time defect prediction,'' \emph{Information and Software Technology}, vol.~87, pp. 206--220, 2017.

\bibitem{24}
M.~K. Thota, F.~H. Shajin, P.~Rajesh \emph{et~al.}, ``Survey on software defect prediction techniques,'' \emph{International Journal of Applied Science and Engineering}, vol.~17, no.~4, pp. 331--344, 2020.

\bibitem{25}
B.~Eken, ``Assessing personalized software defect predictors,'' in \emph{The 40th International Conference on Software Engineering}, 2018, pp. 488--491.

\bibitem{26}
T.~Jiang, L.~Tan, and S.~Kim, ``Personalized defect prediction,'' in \emph{The 28th International Conference on Automated Software Engineering}.\hskip 1em plus 0.5em minus 0.4em\relax Ieee, 2013, pp. 279--289.

\bibitem{13}
J.~Jiarpakdee, C.~Tantithamthavorn, A.~Ihara, and K.~Matsumoto, ``A study of redundant metrics in defect prediction datasets,'' in \emph{The 27th International Symposium on Software Reliability Engineering Workshops}.\hskip 1em plus 0.5em minus 0.4em\relax IEEE, 2016, pp. 51--52.

\bibitem{14}
J.~Jiarpakdee, C.~Tantithamthavorn, and A.~E. Hassan, ``The impact of correlated metrics on defect models,'' \emph{arXiv preprint arXiv:1801.10271}, 2018.

\bibitem{27}
X.~Yang, D.~Lo, X.~Xia, Y.~Zhang, and J.~Sun, ``Deep learning for just-in-time defect prediction,'' in \emph{International conference on software quality, reliability and security}.\hskip 1em plus 0.5em minus 0.4em\relax IEEE, 2015, pp. 17--26.

\bibitem{28}
T.~Hoang, H.~K. Dam, Y.~Kamei, D.~Lo, and N.~Ubayashi, ``Deepjit: an end-to-end deep learning framework for just-in-time defect prediction,'' in \emph{The 16th IEEE/ACM International Conference on Mining Software Repositories}.\hskip 1em plus 0.5em minus 0.4em\relax IEEE, 2019, pp. 34--45.

\bibitem{30}
J.~Li, P.~He, J.~Zhu, and M.~R. Lyu, ``Software defect prediction via convolutional neural network,'' in \emph{International conference on software quality, reliability and security}.\hskip 1em plus 0.5em minus 0.4em\relax IEEE, 2017, pp. 318--328.

\bibitem{29}
L.~Qiao, X.~Li, Q.~Umer, and P.~Guo, ``Deep learning based software defect prediction,'' \emph{Neurocomputing}, vol. 385, pp. 100--110, 2020.

\bibitem{31}
M.~N.~R. Chowdhury, W.~Zhang, and T.~Akilan, ``Anova-based automatic attribute selection and a predictive model for heart disease prognosis,'' \emph{arXiv preprint arXiv:2208.00296}, 2022.

\bibitem{32}
P.~Wei, Z.~Lu, and J.~Song, ``Variable importance analysis: A comprehensive review,'' \emph{Reliability Engineering \& System Safety}, vol. 142, pp. 399--432, 2015.

\bibitem{34}
A.~Gosiewska and P.~Biecek, ``Do not trust additive explanations,'' \emph{arXiv preprint arXiv:1903.11420}, 2019.

\bibitem{36}
J.~Yosinski, J.~Clune, A.~Nguyen, T.~Fuchs, and H.~Lipson, ``Understanding neural networks through deep visualization,'' \emph{arXiv preprint arXiv:1506.06579}, 2015.

\bibitem{37}
D.~Alvarez-Melis and T.~S. Jaakkola, ``On the robustness of interpretability methods,'' \emph{arXiv preprint arXiv:1806.08049}, 2018.

\bibitem{38}
J.~Gou, B.~Yu, S.~J. Maybank, and D.~Tao, ``Knowledge distillation: A survey,'' \emph{International Journal of Computer Vision}, vol. 129, no.~6, pp. 1789--1819, 2021.

\bibitem{39}
G.~Hinton, O.~Vinyals, and J.~Dean, ``Distilling the knowledge in a neural network,'' \emph{arXiv preprint arXiv:1503.02531}, 2015.

\bibitem{che2016interpretable}
Z.~Che, S.~Purushotham, R.~Khemani, and Y.~Liu, ``Interpretable deep models for icu outcome prediction,'' in \emph{AMIA annual symposium proceedings}, vol. 2016, 2017, p. 371.

\bibitem{biggs2021model}
M.~Biggs, W.~Sun, and M.~Ettl, ``Model distillation for revenue optimization: Interpretable personalized pricing,'' in \emph{International conference on machine learning}.\hskip 1em plus 0.5em minus 0.4em\relax PMLR, 2021, pp. 946--956.

\bibitem{43}
M.~Jureczko and L.~Madeyski, ``Towards identifying software project clusters with regard to defect prediction,'' in \emph{The 6th international conference on predictive models in software engineering}, 2010, pp. 1--10.

\bibitem{44}
M.-H. Tang, M.-H. Kao, and M.-H. Chen, ``An empirical study on object-oriented metrics,'' in \emph{The 6th international software metrics symposium}.\hskip 1em plus 0.5em minus 0.4em\relax IEEE, 1999, pp. 242--249.

\bibitem{45}
L.~Qiao and Y.~Wang, ``Effort-aware and just-in-time defect prediction with neural network,'' \emph{PloS one}, vol.~14, no.~2, p. e0211359, 2019.

\bibitem{46}
K.~Zhu, N.~Zhang, S.~Ying, and D.~Zhu, ``Within-project and cross-project just-in-time defect prediction based on denoising autoencoder and convolutional neural network,'' \emph{IET Software}, vol.~14, no.~3, pp. 185--195, 2020.

\bibitem{sivaprasad2023evaluation}
A.~Sivaprasad, E.~Reiter, N.~Tintarev, and N.~Oren, ``Evaluation of human-understandability of global model explanations using decision tree,'' in \emph{European Conference on Artificial Intelligence}.\hskip 1em plus 0.5em minus 0.4em\relax Springer, 2023, pp. 43--65.

\bibitem{parekh2021framework}
J.~Parekh, P.~Mozharovskyi, and F.~d'Alch{\'e} Buc, ``A framework to learn with interpretation,'' \emph{Advances in Neural Information Processing Systems}, vol.~34, pp. 24\,273--24\,285, 2021.

\bibitem{lakkaraju2020robust}
H.~Lakkaraju, N.~Arsov, and O.~Bastani, ``Robust and stable black box explanations,'' in \emph{International conference on machine learning}.\hskip 1em plus 0.5em minus 0.4em\relax PMLR, 2020, pp. 5628--5638.

\bibitem{bang2021explaining}
S.~Bang, P.~Xie, H.~Lee, W.~Wu, and E.~Xing, ``Explaining a black-box by using a deep variational information bottleneck approach,'' in \emph{Proceedings of the AAAI conference on artificial intelligence}, vol.~35, no.~13, 2021, pp. 11\,396--11\,404.

\bibitem{60}
R.~Saleem, B.~Yuan, F.~Kurugollu, A.~Anjum, and L.~Liu, ``Explaining deep neural networks: A survey on the global interpretation methods,'' \emph{Neurocomputing}, vol. 513, pp. 165--180, 2022.

\bibitem{covert2020understanding}
I.~Covert, S.~M. Lundberg, and S.-I. Lee, ``Understanding global feature contributions with additive importance measures,'' \emph{Advances in Neural Information Processing Systems}, vol.~33, pp. 17\,212--17\,223, 2020.

\bibitem{61}
\BIBentryALTinterwordspacing
S.~Bassan, G.~Amir, and G.~Katz, ``Local vs. global interpretability: A computational complexity perspective,'' in \emph{Forty-first International Conference on Machine Learning}, 2024. [Online]. Available: \url{https://openreview.net/forum?id=veEjiN2w9F}
\BIBentrySTDinterwordspacing

\bibitem{51}
J.~Nam, S.~J. Pan, and S.~Kim, ``Transfer defect learning,'' in \emph{The 35th international conference on software engineering}.\hskip 1em plus 0.5em minus 0.4em\relax IEEE, 2013, pp. 382--391.

\bibitem{52}
R.~Wu, H.~Zhang, S.~Kim, and S.-C. Cheung, ``Relink: recovering links between bugs and changes,'' in \emph{The 13th European conference on Foundations of software engineering}, 2011, pp. 15--25.

\bibitem{53}
M.~D’Ambros, M.~Lanza, and R.~Robbes, ``Evaluating defect prediction approaches: a benchmark and an extensive comparison,'' \emph{Empirical Software Engineering}, vol.~17, pp. 531--577, 2012.

\bibitem{54}
F.~Peters and T.~Menzies, ``Privacy and utility for defect prediction: Experiments with morph,'' in \emph{The 34th International conference on software engineering}.\hskip 1em plus 0.5em minus 0.4em\relax IEEE, 2012, pp. 189--199.

\bibitem{18}
J.~Jiarpakdee, C.~Tantithamthavorn, and C.~Treude, ``Autospearman: Automatically mitigating correlated software metrics for interpreting defect models,'' in \emph{The 34th International Conference on Software Maintenance and Evolution}, 2018, pp. 23--29.

\bibitem{RUS}
S.~Feng, J.~Keung, Y.~Xiao, P.~Zhang, X.~Yu, and X.~Cao, ``Improving the undersampling technique by optimizing the termination condition for software defect prediction,'' \emph{Expert Systems with Applications}, vol. 235, p. 121084, 2024.

\end{thebibliography}

\end{document}